\documentclass[referee,a4paper,12pt,traditabstract]{jswsc} 

\usepackage{graphicx}
\usepackage{txfonts}
\usepackage{subfigure}
\usepackage{epstopdf}
\usepackage[mathlines]{lineno}
\usepackage[authoryear,round]{natbib}
\usepackage[backref]{hyperref}
\makeatletter
\AtBeginDocument{%
    \catcode`_=12
    \begingroup\lccode`~=`_
    \lowercase{\endgroup\let~}\sb
    \mathcode`_="8000
    \immediate\write\@auxout{\catcode`_=12 }%
    \immediate\write\@auxout{\catcode`^=12 }%
}
\makeatother
\usepackage{url}

\hypersetup{colorlinks=true,citecolor=cyan,urlcolor=cyan,linkcolor=blue}

\begin{document}


   \title{P2D - A Two-Dimensional Multi-Spacecraft Solar Wind Persistence Model}

   
   \titlerunning{2D Solar Wind Persistence Model}

   \authorrunning{Milosic et al.}

   \author{D. Milošić
          \inst{1} 
          \and
          M.Temmer
          \inst{1}
                    \and
          S. G. Heinemann
          \inst{2}
                    \and
          M.Owens
          \inst{3}
                    \and
          S. J. Hofmeister
          \inst{4}
          }

   \institute{Institute of Physics, University of Graz, Austria\\
              \email{\href{mailto:danielmilosic@live.com}{danielmilosic@live.com}}
         \and
             Department of Physics, University of Helsinki, Finland \\
          \and
             Department of Meteorology, University of Reading, UK \\
        \and
             Columbia Astrophysics Laboratory, Columbia University, 538 West 120th Street, New York, NY 10027, USA
             }


 
  \abstract
 {The background solar wind is a key component in space weather forecasting, as it contains geoeffective high-speed streams and provides the medium through which coronal mass ejections propagate and evolve in interplanetary space. Due to the solar rotation and the slow evolution of large-scale solar wind structures, solar wind properties exhibit an autocorrelation with a period of roughly 27 days, particularly at solar minimum. We made use of this property to develop a solar wind persistence model with input from multiple spacecraft (Solar Orbiter, Parker Solar Probe, STEREO-A, STEREO-B and the OMNI database). The model ballistically propagates in-situ data from the position of their measurement radially away from the Sun, as well as longitudinally with the solar rotation rate, producing 2D maps of solar wind parameters. These can be extracted at any point in the heliosphere for a solar wind reconstruction. From Earth’s perspective, the reconstruction performs with an MAE of 65.41 km~s$^{-1}$ and 3.51 $cm^{-3}$ for speed and density, respectively. During high-speed streams, the peak hit rate is 50\% and 48\% for speed and density. It performs best during times when there are spacecraft located between Earth and Lagrange point L5. During these times, the mean absolute error of the predicted solar wind speed decreases by roughly 35\% in comparison to the benchmark 27-day persistence model.  Therefore, future L5 missions like Vigil are expected to provide a robust basis for reliable persistence forecasts. 
 
   }        





   \maketitle
\section{Introduction}


The solar wind is a continuous stream of ionized plasma which fills interplanetary space. It can be generally divided into two modes: fast wind with speeds $>$ 450 km~s$^{-1}$ and slow wind with speeds $<$ 450 km~s$^{-1}$ \citep[see e.g.,][]{Wolfe1966}. As the fast solar wind encounters slow streams, plasma is compressed at their intersection, resulting in enhanced plasma density and magnetic field \citep{BelcherDavis1971}. Due to the solar rotation, these interaction regions form a spiral structure as the plasma propagates radially away from the Sun \citep{BelcherDavis1971, Pizzo1978}, i.e. forming the Parker Spiral \citep{Parker1958}. The steady outflow of solar wind makes them appear to corotate with Sun, which is why we refer to them as corotating interaction regions \citep[CIRs,][]{Tsurutani2006}. CIRs cause large, periodical variations in plasma and magnetic field profiles of the solar wind. 

The source regions of the fast solar wind are coronal holes \citep[CH;][]{Krieger1973,Nolte76}, which are stable and long-lived structures in the solar corona \citep[see, e.g.,][]{Sargent1985}. The lifetimes of CIRs can therefore reach up to more than 10 Carrington rotations \citep{Temmer2007, Heinemann2018, Heinemann2020}, leading to a slow evolution of the heliospheric plasma environment, especially during solar minimum. This simplifies the task of forecasting ambient solar wind conditions, which has multiple applications in space weather prediction.

While CIRs themselves are of direct space-weather interest \citep{Alves2006, TemmerReview2021}, a better understanding of the background solar wind conditions is identified to be crucial for improving forecasts of the kinematics of coronal mass ejections \citep[CMEs; see recommendations in][]{Temmer2023}. CMEs are magnetized plasma ejections originating from the Sun, that can cause a multitude of effects on Earth, ranging from satellite orbits decaying, to aurorae and ground induced currents \citep[see, e.g.,][]{Schrijver2014}. Their path from the Sun to Earth is strongly impacted by the properties of the ambient solar wind, affecting the speed as well as the morphology of CMEs \citep{Gopalswamy2000, Savani2010, Temmer2011, Heinemann2019, Davies2021, Remeshan2026}. In that respect, the drag force acts as a primary mechanism, where speed and density differences between the CME and ambient solar wind are the most critical factors \citep[e.g.,][]{Vrsnak2001,Cargill2004}. Therefore, correctly predicting the speed and density profiles of the solar wind from Sun to Earth has the potential to improve arrival time and speed predictions of CMEs \citep[e.g.,][]{Case2008}.

Multiple approaches to solar wind modeling exist, ranging from empirical approaches to fully physics-based magnetohydrodynamic simulations, with increasing physical complexity and computational cost. The simplest and most computationally cheap models consist of algorithms based on empirical relations. The empirical solar wind forecast \citep[ESWF,][]{ESWF2015, ESWF2016, ESWF2023} exploits the linear relationship between CH areas and solar wind speed, forecasting solar wind speed with roughly 4 days of lead time \citep{ESWF2007}. The Wang-Sheeley-Arge (WSA) model \citep{Arge2000} estimates solar wind speed based on the magnetic expansion factor. To account for the interaction of solar wind streams traveling at different speeds through interplanetary space, the model employs a simplified interaction scheme that adjusts their relative velocities.

Persistence models \citep{Owens2013, Temmer2018} use the slow evolution of large-scale solar wind conditions and assume that the in-situ measured profile persists over one Carrington rotation. \cite{Owens2013} showed that solar wind parameters exhibit an autocorrelation with a period of equal to the average solar rotation rate. This autocorrelation persists over multiple solar rotations, decaying with time. In \cite{Milosic2026}, we investigated this effect with multiple spacecraft scattered across the heliosphere, showing that the persistence of the solar wind decays exponentially over time with a characteristic decay time of roughly 50 days. \cite{Turner2021} find that persistence models are reliable as long the latitudinal offset between spacecraft stays below 5$^\circ$. Persistence models currently consistently outperform Magneto-hydrodynamic (MHD) models and are often used as a benchmark for model performance \citep{Reiss2016, MacNeice2018, Temmer2018, Heinemann2025}.

Introducing more physical processes, hydrodynamical models \citep[e.g. HUXt,][]{HUXt} neglect the magnetic characteristics (as well as forces from gravity or pressure gradients) of the solar wind and approximate it as a fluid, obeying hydrodynamic laws. \cite{Owens2026} use the same hydrodynamic model, but driven by in-situ observations. The in-situ observations from 1AU are backmapped to a source surface at 0.1AU and used as input to the HUXt model, avoiding the reliance on magnetic field maps of the solar photosphere. Similarly, \cite{Turner2025} have shown that using data assimilation \citep{bravda2019} with near-Earth in-situ measurements can significantly improve the HUXt's representation of the ambient solar wind. 

MHD models solve the entire set of MHD equations in one to three dimensions to produce a forecast \citep[see, e.g.,][]{Odstrcil2003, SUSANOO, EUHFORIA}. Recently, MHD models have been developed, using time-dependent boundary conditions for dynamically evolving results \citep{Linker2016, Merkin2016, GAMERA, ICARUS, Owens2024, Samara2025}.

Machine learning–based models rely on large training datasets and are computationally expensive during training, though typically inexpensive at inference. So far, they offer single-point forecasts at Earth, rather than global solutions \citep[see, e.g.,][]{Bailey2021, Collin2025, Abraham2026, Billcliff2026}. 

Overall, solar wind forecasting models trade physical complexity for computational efficiency, with ongoing efforts focused on bridging this gap through data assimilation, improved boundary conditions, and hybrid approaches. In this work, we exploit the abundance of in-situ measurements at different positions in the heliosphere giving us realistic boundary conditions. We are therefore able to neglect complex MHD processes in the solar wind, relying solely on (1) the assumption of persistence and (2) a simple inelastic collision scheme for fast/slow solar wind interactions.  

In this article, we present a novel approach to persistence modeling which we call Persistence model in 2 Dimensions (P2D). Based on previous studies by \cite{Milosic2025} and \cite{Milosic2026}, we use in-situ solar wind speed and density measurements from multiple spacecraft at different heliospheric positions, from which we construct a two-dimensional map covering most of the inner heliosphere. This two-dimensional map is then used to generate solar wind speed and density reconstructions. In this work we extract reconstructions at Earth location, which are validated against OMNI data and compared to existing ambient solar wind models using standard performance metrics.

\section{Methods}

This section describes the data and algorithm used to produce the P2D solar wind reconstruction.

\subsection{Data}

\begin{figure}
   \centering
   \subfigure{\includegraphics[width=0.9\columnwidth]{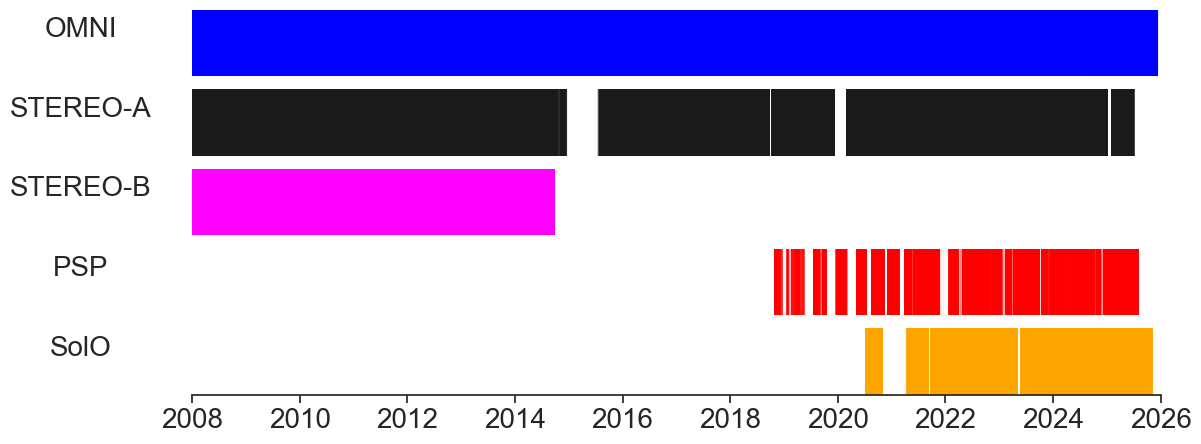}}

   \caption{\label{fig:Data} \small Data availability of all involved spacecraft.} 

   \end{figure}

As input, we use science-level data of the bulk solar wind speed and density as measured by spacecraft at five different locations in the heliosphere.  Figure \ref{fig:Data} shows the data availability of all five sources over time.

OMNI \citep{OMNI} is a database providing solar wind data measured at L1 by the Advanced Composition Explorer \citep[ACE,][]{ACE} and the Wind \citep{WindSWE} spacecraft. The data are numerically propagated to a modeled position of Earth's bowshock, and hourly averaged. The time shift is being performed using the in-situ measured velocity vector and the plane of the phase front \citep{OMNI}. We use its data for the entire investigated time range, from 1 February, 2007 to 31 December, 2025 with only a few small data gaps.

The Solar TErrestrial RElations Observatory \citep[STEREO,][]{STEREO} consists of two spacecraft, both roughly at 1AU distance to the Sun; Ahead (STEREO-A) and Behind (STEREO-B). STEREO-A has a slightly faster orbital speed than the Earth and STEREO-B slightly slower, making the spacecraft separate from each other by about 44$^\circ$ in longitude per year. Their PLAsma and SupraThermal Ion Composition \citep[PLASTIC,][]{PLASTIC} instruments provide us with proton density and speed measurements. We use STEREO-A data for the entire time range, while STEREO-B data only extends up until 1 October, 2014, when contact to the spacecraft was lost. Other than that, there are only a few data gaps, as can be seen in Figure \ref{fig:Data}. We do not fill data gaps artificially.

Parker Solar Probe \citep[PSP,][]{PSP} with its Solar Wind Electrons Alphas and Protons \citep[SWEAP,][]{PSPSWEAP} instrument provides data since its launch on 12 August, 2018 up until 31 July, 2025. PSP has a highly elliptical orbit, reaching distances as close as 9 solar radii to the Sun. Due to memory and bandwidth restrictions, the instrument is not switched on the entire time, resulting in significant data gaps, as can be seen in Figure \ref{fig:Data}. In an operational setting, PSP data would not be used due to its large data-latency.

Solar Orbiter \citep[SolO,][]{SolO} also has an elliptical orbit, with distances between 0.29AU and 1AU.  Its Solar Wind Analyser \citep[SWA][]{SolOSWA} instrument provides data between 15 June, 2020 and 31 October, 2025 with some data gaps, as can be seen in Figure \ref{fig:Data}. 

The spacecraft and their instruments do not individually cover the entire time range, but have large overlaps. We note that some instruments are not switched on during extended periods of their mission durations, leading to data gaps, which we do not fill artificially.

\subsection{2D Propagation}

The basis for the P2D model is an assumption of a persisting solar wind. The algorithm follows the exact same principles as laid out in prior work by \cite{Milosic2025}. We assume that in-situ measurements of speed and density remain unchanged, while propagating across two dimensions: longitude and radial distance from the Sun. We assume (a) all flows to be radial, (b) all spacecraft to be at the ecliptic, and (c) the solar wind to rotate rigidly with a synodic period of 27.27 days. Assumption (a) is reasonable as the radial solar wind velocity component is typically the largest by far. Assumption (b) holds for the majority of the dataset, but SolO orbits with an inclination, where there is a latitudinal difference of up to 24$^{\circ}$ during the investigated times. Assumption (c) is true on average, but there are slight variations in the corotational speed of CIRs \citep{Allen2020}. Assumption (c) is also not valid for ICMEs.

In our simulation the in-situ measured data are being propagated (1) across heliographic longitude with the average solar rotation of 27.27 days and (2) across radial distance from the Sun with a propagation speed equal to their in-situ measured proton speed. After a reduction of each spacecraft's data set to a resolution of $0.5^\circ$ in longitude (i.e. 54-minute time-resolution for OMNI), each data point is being (1) moved longitudinally. Meanwhile, the same data point is (2) being copied and propagated radially with the speed that it has according to the measurement. As all data points propagate longitudinally, their copies produce spiral lines with their radial movement. If data points approach each other within a minimal threshold distance, they interact by an ideal inelastic collision (see Equation \ref{Eq:inelastic}). In an intermediate step (2.5), the resulting maps from all available spacecraft are combined. Profiles from different spacecraft are treated separately, so that no two data points with origins in different spacecraft can interact. (3) Solar wind speed and density values at the Earth’s position, or any other target in the inner heliosphere, are extracted from the two-dimensional map constructed from each individual spacecraft’s data, and (4) according to the error statistics \citep[see Equations \ref{eq:MAE_V} provided by][and \ref{eq:MAE_N}]{Milosic2026}, these are combined to give a single, one-dimensional solar wind speed and density reconstruction at Earth including uncertainties.

\begin{figure}
   \centering
   \subfigure{\includegraphics[width=0.9\columnwidth]{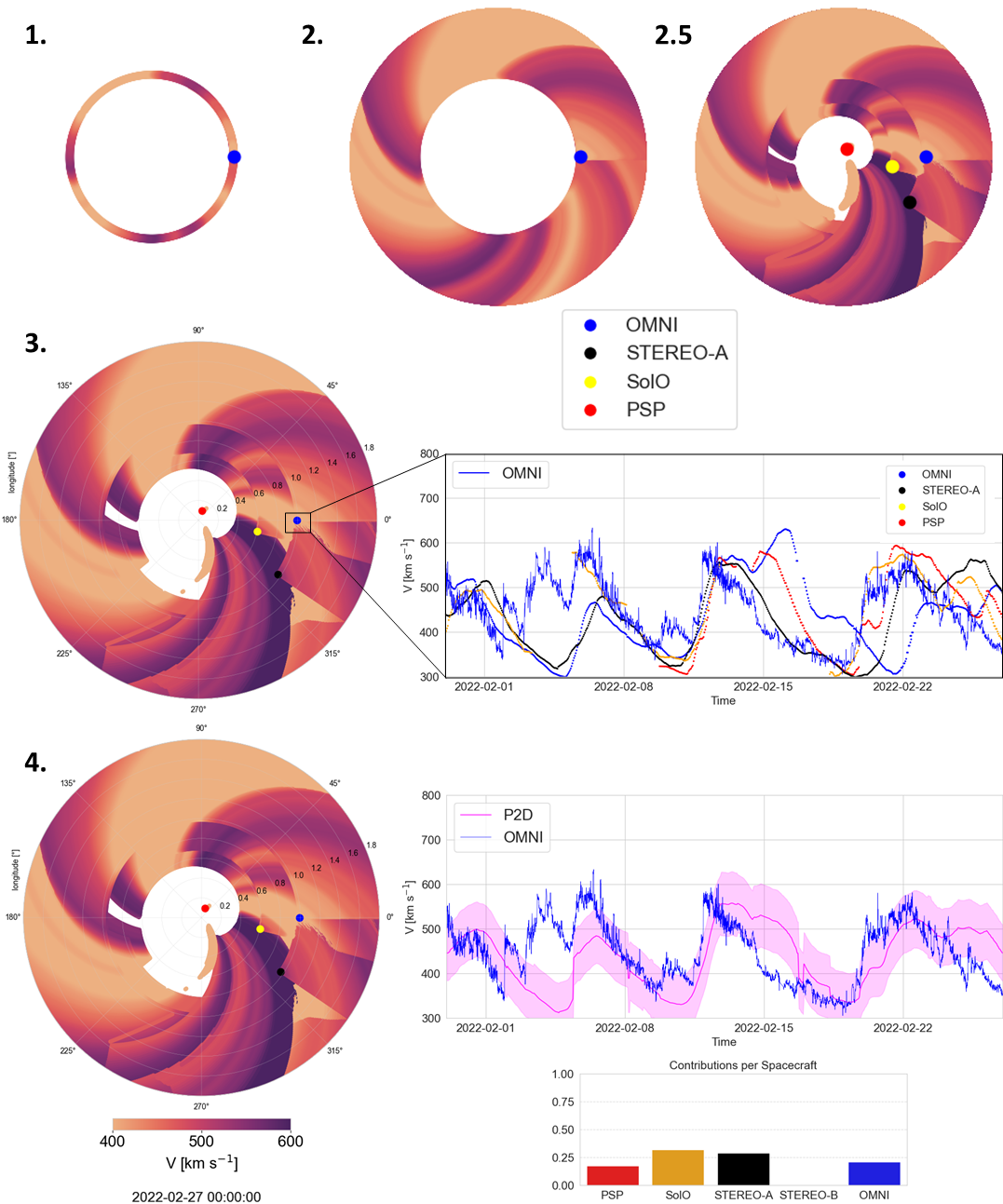}}

   \caption{\label{fig:2DMaps} \small The 4 steps of the P2D workflow. Each polar plot shows a two-dimensional map of solar wind speed with the big colored dots representing different spacecraft. Panel 1: OMNI data (blue) being propagated in longitude (HEEQ) for one Carrington rotation. Panel 2: OMNI data being propagated across longitude (HEEQ) and radial distance. Panel 2.5: OMNI (blue), STEREO-A (black), PSP (red) and SolO data (yellow) being propagated across longitude (HEEQ) and radial distance for one Carrington rotation. Panel 3: The left side is identical to Panel 2.5 and the right side shows with the same color code as before the extracted Time series from this simulation at Earth position. Panel 4: Same as Panel 3, but the right side now shows the combined, one-dimensional reconstruction, stitched together from the individual profiles in Panel 3. The plot on the lower right side of Panel 4 shows the relative contribution from each source to the reconstruction. The accompanying video shows the same steps dynamically, starting from a video of Panel 1 and ending with video of Panel 4, lasting a total of 1 minute and 15 seconds.} 

   \end{figure}

Figure \ref{fig:2DMaps} and its corresponding animation summarize the algorithm in four steps, which are described in more detail below.

\subsubsection{Longitudinal propagation (Step 1)}
\label{sec:longitudinal}

P2D propagates data, which are measured in-situ at each given location, across heliographic longitude with the average solar rotation of 27.27 days over 360$^\circ$. The resolution can strongly impact the performance metrics of a model \citep{Riley2013, Majumdar2025}. In the interest of comparability to other models, we chose a longitudinal resolution of 0.5°, which corresponds to slightly less than an hour (54min) at 1AU. This cadence is different for each spacecraft due to their orbital speeds. Panel 1 in Figure \ref{fig:2DMaps} shows OMNI data being propagated longitudinally for one Carrington rotation for visualization purposes. In the Heliocentric Earth Equatorial (HEEQ) coordinate system, the resulting structure is a 360$^\circ$ ring around the Sun, color coded with the bulk speed, based on measurements provided by OMNI. This process is repeated for every available spacecraft.

Since we do not filter out ICMEs, during this step, in-situ measured ICMEs are also being propagated across longitude. We do this in the interest of comparability to other models, but in an operational setting, ICMEs should be removed, as will be discussed in Sections \ref{sec:Metrics} and \ref{sec:Discussion}.

\subsubsection{Radial propagation (Step 2)}
\label{sec:radial}

In the next step, shown in Panel 2 of Figure \ref{fig:2DMaps}, each data point is being propagated radially away from the Sun. Assuming constant speed, the propagation speed is defined by the in-situ measured bulk speed. If a data point reaches the vicinity ($\frac{1}{2}$ of the model resolution) of another, the two collide inelastically and they move on with their average momentum, similar to the WSA Model \citep{Arge2000}. This prevents different streams overtaking each other, which would result in nonphysical results and reconstructions. The process of inelastic collision can be described as:

\begin{equation}\label{Eq:inelastic}
    V_i(t+1) = V_j(t+1) = \frac{V_i (t) \cdot N_i (t) + V_j(t)\cdot N_j(t)}{N_i(t) +  N_j(t)},
\end{equation}

where V is the bulk speed, N is the proton density, t is the time step and i, j are different data points \citep[see also][]{Milosic2026}. This is performed for each spacecraft.

Panel 2.5 shows an intermediate step of combining the maps from all available spacecraft. The maps are produced simultaneously; we only show this as an intermediate step for better understanding of the workflow and visualization. Crucially, no two data points from different spacecraft interact with each other through the inelastic collision scheme or otherwise.

\subsubsection{Extraction of model results at Earth (Step 3)}

Panel 3 of Figure \ref{fig:2DMaps} illustrates the extraction of time series from the 2D map produced in the previous step (cf.\,Panel 2.5). Time series are extracted for the Earth location from each source -- STEREO-A (black dotted line), SolO (yellow dotted line), PSP (red dotted line), and OMNI data from the previous Carrington rotation (blue dotted line) -- and compared against OMNI in-situ measurements (blue solid line). Data points are virtually sampled as they reach the vicinity of Earth, which is defined by a threshold of 0.01 AU, multiplied by the data resolution in degrees (0.5). Out of those data points, only the one that is located closest to Earth is kept for the reconstruction. This process occurs at every model timestep for each in-situ source, resulting in the simultaneous generation of multiple time series.

\subsubsection{Composite reconstruction from multiple spacecraft (Step 4)}
\label{sec:Combination_to_1D}

To produce a speed and density reconstruction at Earth, we combine the profiles from all different sources into a single profile, as shown by Panel 4 in Figure \ref{fig:2DMaps}. To this end, we combine the profiles based on different spacecraft linearly, with weights appropriate to the spacecraft's position in the heliosphere. The weighting of the contribution per spacecraft depends on the longitudinal and latitudinal spacecraft separation to Earth using statistical results from \cite{Milosic2026}. They found that the mean absolute error (which we use as an estimate for the prediction error $\Delta V$ of the speed) in a persistence simulation can be calculated using the heliographic latitudinal and longitudinal spacecraft separation:

\begin{equation}
\label{eq:MAE_V}
    \Delta V(\Delta \lambda, \Delta \phi) = 2.4\cdot\Delta \lambda + 9.1\cdot ln(\Delta \phi) + 17.4 ,
\end{equation}

where $\Delta \lambda$ is the heliographic latitudinal separation in degrees and $\Delta \phi$ is the longitudinal separation. As the plasma density is also relevant for space weather purposes (e.g., for calculating the drag force or ram pressure), we repeated the analysis from \cite{Milosic2026} to get an estimate of the density errors. Figure \ref{fig:N_COR} shows the same analysis as performed in \cite{Milosic2026}, but for density instead of speed. Here, we investigate trends of the Mean Absolute Error (MAE) across (a) latitudinal separation, (b) longitudinal separation and (c) radial separation. We find similar trends for the density as for the speed, with linear effect of heliographic latitude, logarithmic dependence on longitude, and no trend for radial distance. 

From this analysis we receive an estimate for the density error $\Delta N$:

\begin{equation}
\label{eq:MAE_N}
    \Delta N (\Delta \lambda, \Delta \phi) = 0.03\cdot\Delta \lambda + 0.23\cdot ln(\Delta \phi) +2.0 .
\end{equation}

We take the error estimates for speed and density as an uncertainty measure for our reconstruction. Additionally, of each parameter the inverse of the uncertainty is used as weight for the linear combination of different spacecraft measurements.

Each spacecraft s contributes an amount in proportion to the inverse of the associated uncertainty:

\begin{equation}
\label{eq:Contributions}
    C_{s, X} = \frac{\frac{1}{\Delta X_{s}}}{\sum_i{\frac{1}{\Delta X_{i}}}},
\end{equation}

where $\Delta X_i$ is the uncertainty of the reconstruction for parameter $X$ based on data provided by spacecraft $i$. The total reconstruction uncertainty is then the sum of all contributions, multiplied by their respective individual uncertainty:

\begin{equation}
\label{eq:Combination}
    \Delta X = {\sum_s{C_{s, X} \cdot \Delta X_{s}}}
\end{equation}

   \begin{figure}
   \centering
   \subfigure{{\includegraphics[width=\columnwidth]{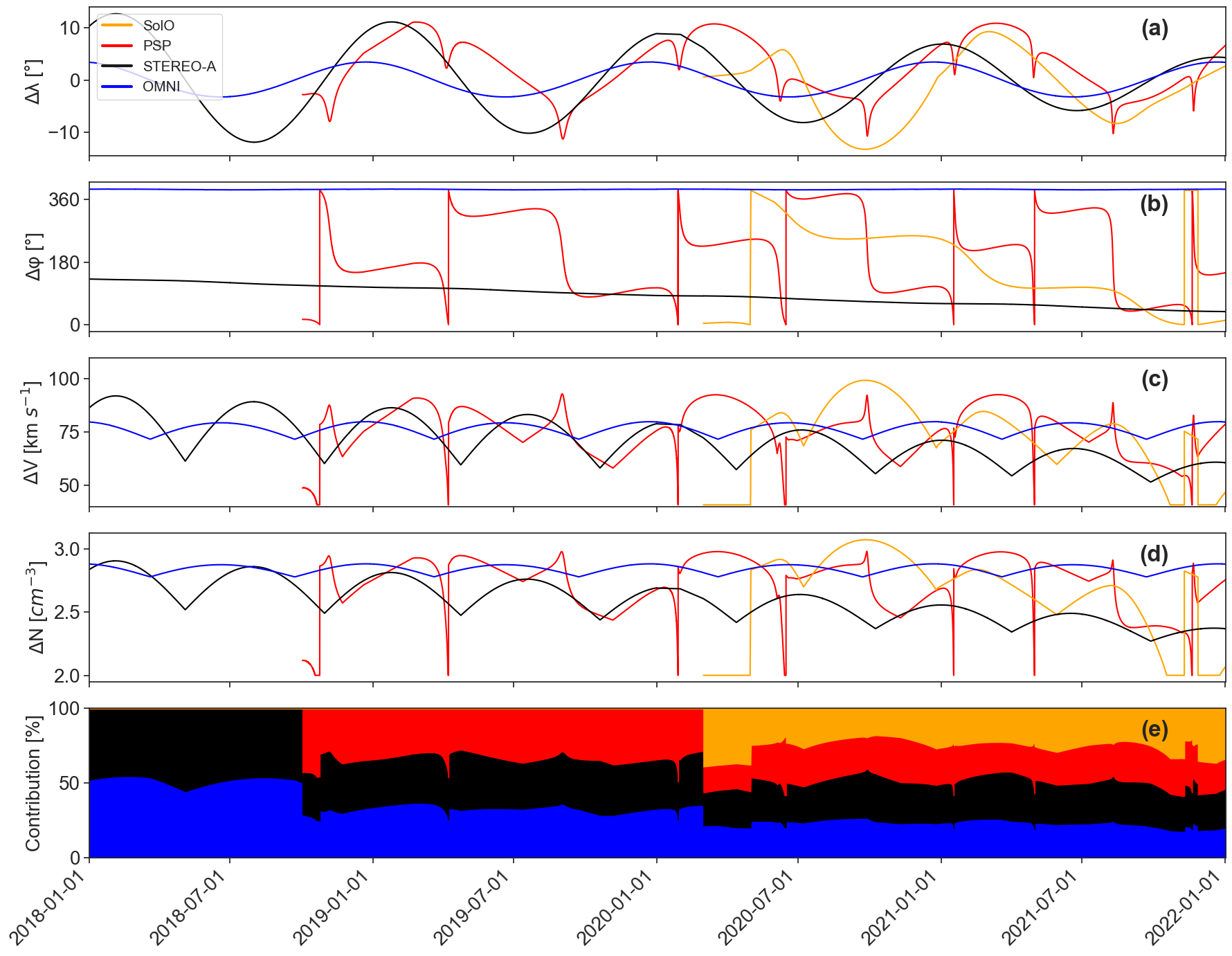}}}

   \caption{\label{fig:Spaghetti}\small Spacecraft orbits and resulting reconstruction uncertainties and contributions. Yellow: SolO; Red: PSP; Black: STEREO-A; Blue: OMNI. Panel (a): Heliographic latitudinal separations to Earth; (b): Longitudinal separations to Earth; (c): Speed uncertainty; (d): Density uncertainty; (e): Relative contributions of each spacecraft to the reconstruction.} 
   \end{figure}

Figure \ref{fig:Spaghetti} shows how the spacecraft orbits affect the combination of time series into the near-Earth prediction. Panels (a) and (b) show the heliographic latitudinal and longitudinal separation of the input spacecraft to Earth. Note that the longitudinal separation exceeds 360$^\circ$ in HEEQ, due to Earth's longitudinal motion during a Carrington rotation. Panels (c) and (d) show the speed and density uncertainties, resulting directly from evaluating the data from Panels (a) and (b) with Equations \ref{eq:MAE_V} and \ref{eq:MAE_N}. The bottom Panel, (e), shows the individual spacecraft contributions to the speed prediction as a percentage. Figure \ref{fig:Spaghetti_long} in the Appendix shows the same panels for the entire time range.

The launches of PSP (August 12, 2018) and SolO (February 10, 2020) are marked by their sudden appearance as the red and yellow profiles in each of the panels. Right after their respective launches, we can see in Panel (e) that their contributions are very large, due to the small longitudinal and latitudinal separations from Earth after the start of the missions. Accordingly, Panels (c) and (d) show minimal speed and density uncertainties. In Panel (a), we can also identify the PSP perihelia by looking at the sharp dips in PSP latitude. During those times, PSP's contribution in Panel (e) also sharply changes, either to larger or smaller contributions, depending on absolute latitudinal difference to Earth.

\subsection{Resulting Products}

\begin{figure}
   \centering
   \subfigure{{\includegraphics[width=\columnwidth]{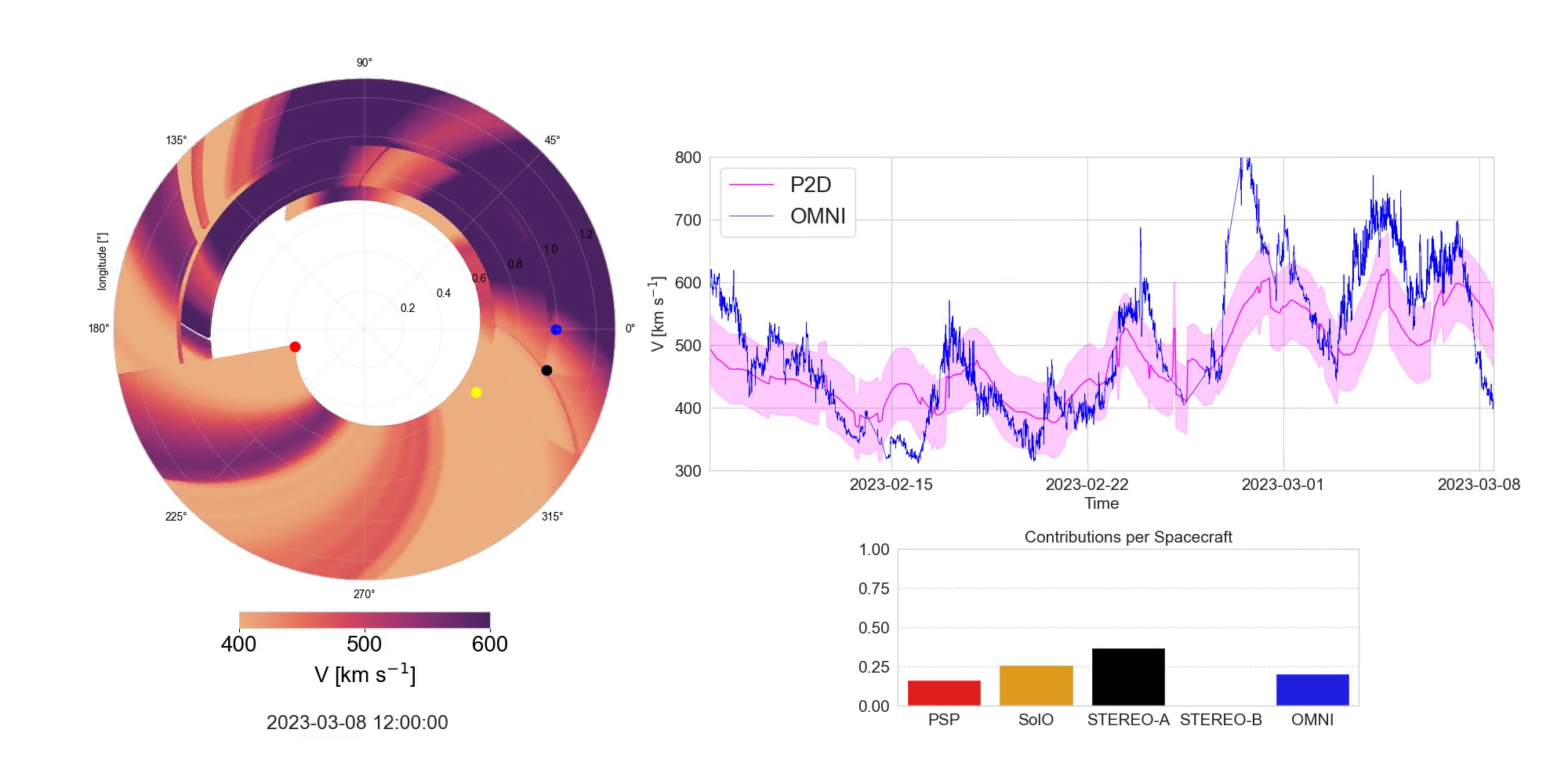}}}

   \caption{\label{fig:TS}\small Final two-dimensional speed map and time series extracted from the map at Earth location on March 6, 2023, 12:00 UT. Left: 2D map, color-coded with the reconstructed solar wind bulk speed. Upper right: Time series of the solar wind bulk speed at Earth from the OMNI database in blue and P2D reconstruction in magenta with a light magenta uncertainty area. Lower right: Relative contribution from each spacecraft or database to the reconstruction at the shown time. A movie of the entire reconstructed time range between January 2008 and December 2025 is available in the online version of the article, lasting 14 minutes and 36 seconds.} 
   \end{figure}

\begin{figure}
   \centering
   \subfigure{{\includegraphics[width=\columnwidth]{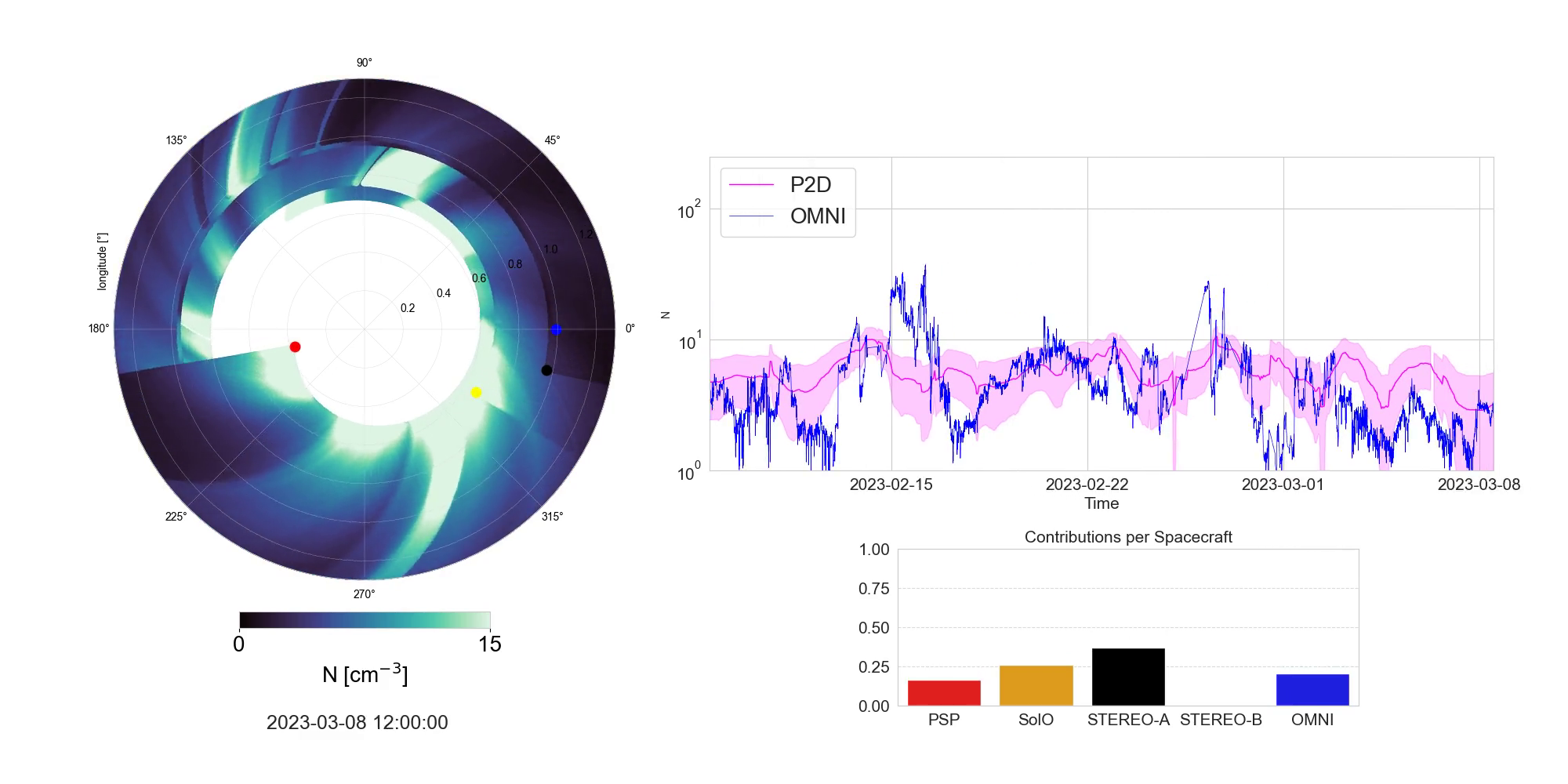}}}

   \caption{\label{fig:TS_N}\small Final two-dimensional density map and time series extracted from the map at Earth location on March 6, 2023, 12:00 UT. Left: 2D map, color-coded with the reconstructed solar wind proton density. Upper right: Time series of the solar wind proton density at Earth from the OMNI database in blue and P2D reconstruction in magenta with a light magenta uncertainty area. Lower right: Relative contribution from each spacecraft or database to the reconstruction at the shown time. A movie of the entire reconstructed time range between January 2008 and December 2025 is available in the online version of the article, lasting 14 minutes and 36 seconds.} 
   \end{figure}

Figures \ref{fig:TS} and \ref{fig:TS_N} and their respective animations show the resulting models after performing all the above mentioned steps. On the left side of each figure, there are two-dimensional maps, showing speed and density, respectively. The maps are model results in their own right and can be output as 2D data, representing the P2D model. On the right side there are time series, showing the evolution of these profiles as they are virtually measured at Earth and combined into a single profile, as described in Section \ref{sec:Combination_to_1D}. The corresponding weights are shown below the respective time series. Note that, under the condition that there are spacecraft closer to the Sun, this kind of reconstruction can be produced at any point in the heliosphere, not just Earth. 
Running the model for one year of spacecraft data of a single spacecraft at 0.5$^\circ$ resolution takes about 45 seconds on a laptop with a 2.6 GHz CPU.
The performance of the resulting one-dimensional reconstructions, both in speed and density, will be assessed in the next sections. 

\section{Performance Metrics}
\label{sec:Metrics}

\subsection{Metrics}

We quantify the performance of the P2D model for the bulk speed and density components at Earth in accordance with the unified scheme provided by \cite{Reiss2023}. Whenever possible, we chose the metrics that are also available at CAMEL\footnote[1]{https://ccmc.gsfc.nasa.gov/camel/AmbientSolarWind/} from the Community Coordinated Modeling Center (CCMC) to make them comparable to other models. 

For bulk speed we present point-to-point metrics, binary metrics as well as a peak analysis, comparing modelled CIR-peak speeds to observed ones. For proton density, we present results of point-to-point metrics as well as a CIR-density-peak analysis. As density models are less frequently available, there are fewer models to compare to.

We set our findings in perspective by comparing the metrics between the P2D model and other available solar wind models, namely the 27-day persistence model \citep{Owens2013}, the Empirical Solar Wind Forecast \citep[ESWF;][]{ESWF2007, ESWF2015, ESWF2016, ESWF2023}, the Heliospheric Upwind eXtrapolation with time dependence model driven by OMNI in-situ data with four days of lead-time \citep[HUXt;][]{HUXt, Owens2026}, the Wang-Sheeley-Arge model 2.2 \citep[WSA 2.2;][]{Arge2000} and the Ambient Solar Wind Prediction using Machine Learning model \citep[AmbSoWi-ML;][]{Bailey2021}. 

Table \ref{tab:metric_definitions} shows the metrics we use as well as their definitions, where $n$ is the amount of data pairs, $X$ are the observables, $\hat{X}$ is the mean of all $X$, TP is the amount of true positives, FP is the amount of false positives, TN is the amount of true negatives and FN is the amount of false negatives.  

\renewcommand{\arraystretch}{1.5}
\begin{table}[]
    \centering
    \begin{tabular}{c|c|c}
         Metric Name & Abbreviation & Definition\\
         \hline
         ME & Mean Error & $\frac{1}{n}\sum_{i=1}^{n} (X_i - \hat{X}_i)$\\
         \hline
         MSE & Mean Square Error & $\frac{1}{n}\sum_{i=1}^{n} (X_i - \hat{X}_i)^2$\\
         \hline
         MAE & Mean Absolute Error & $\frac{1}{n}\sum_{i=1}^{n} |X_i - \hat{X}_i|$\\
         \hline
         RMSE & Root Mean Square Error & $\sqrt{\frac{1}{n}\sum_{i=1}^{n} (X_i - \hat{X}_i)^2}$\\
         \hline
         CC & Pearson Correlation Coefficient &
         $\frac{\sum_{i=1}^{n}(X_i-\bar{X})(\hat{X}_i-\bar{\hat{X}})}
         {\sqrt{\sum_{i=1}^{n}(X_i-\bar{X})^2\sum_{i=1}^{n}(\hat{X}_i-\bar{\hat{X}})^2}}$\\
         \hline
         TPR & True Positive Rate & $\frac{TP}{TP + FN}$\\
         \hline
         FPR & False Positive Rate & $\frac{FP}{FP + TN}$\\
         \hline
         TS & Threat Score & $\frac{TP}{TP + FP + FN}$\\
         \hline
         TSS & True Skill Score & $\frac{TP}{TP + FN} - \frac{FP}{FP + TN}$\\
         \hline
         BS & Bias & $\frac{TP + FP}{TP + FN}$\\
    \end{tabular}
    \caption{Definitions of performance metrics.}
    \label{tab:metric_definitions}
\end{table}
\renewcommand{\arraystretch}{1.0}

\subsection{Results - Bulk Speed}
\subsubsection{Point-to-point Metrics}

To ensure consistency in the point-to-point metrics, we limit our comparison to models that cover a sufficient time range and overlap entirely with one another. This drastically confines the number of models for comparison but ensures the accurate depiction of each model's performance. As ICMEs are not removed from the data on CAMEL's scoreboard, we also don't remove them for both the point-to-point metrics as well as the binary metrics. Due to our constraints, the investigated time period for this analysis is from May 19, 2012 to May 26, 2017. This means that PSP and SolO do not contribute to P2D in the point-to-point metric analysis. 

Table \ref{tab:point-to-point-metrics} shows the performance of the bulk speed prediction of the P2D model in comparison to other ambient solar wind models in terms of point-to-point metrics. The first four columns, P2D, 27-day persistence, ESWF and HUXt show the results of our analysis, while all the other columns present results obtained from CAMEL. The values printed in bold indicate the best performing model for the metric.

P2D ranks second-best in terms of ME, with a value of $-$0.32km~s$^{-1}$, which means that on average the in-situ bulk speed is only very slightly underestimated, whereas the 27-day persistence model performs best with an ME of $-$0.06 km~s$^{-1}$. In all other metrics P2D ranks second-best to the machine learning-based model AmbSoWi-ML. P2D performs with an MSE of 7761.13 km~s$^{-1}$ (AmbSoWi-ML: 6502.55 km~s$^{-1}$), an MAE of 65.41 km~s$^{-1}$ (AmbSoWi-ML: 62.26 km~s$^{-1}$), an RMSE of 88.10 km~s$^{-1}$ (AmbSoWi-ML: 80.64) and a Pearson correlation of 0.51 (AmbSoWi-ML: 0.56) only slightly worse than this sophisticated machine learning model.

\begin{table}[]
    \centering
    \begin{tabular}{c|c|c|c|c|c|c|c}
         Metric & P2D & 27d Persistence & ESWF & HUXt & WSA 2.2$^1$  &  AmbSoWi-ML$^1$\\
         \hline
         ME& $-$0.32 & $-$0.06 & $-$15.65 & $-$2.84 & 20.61  & 3.61\\
         \hline
         MSE& 7761.13 & 9966.03 & 11090.27 & 9303.43 & 11118.1  & 6502.55 \\
         \hline
         MAE& 65.41 & 73.85 & 79.39 & 71.20 &  82.29  & 62.26 \\
         \hline
         RMSE& 88.10 & 99.83 & 105.31 & 96.45 & 105.44  & 80.64 \\
         \hline
         CC& 0.51 & 0.45  & 0.44 & 0.48 & 0.37  & 0.56 \\
         \hline
         Cadence & 1h & 1h & 1h & 23min & 1day & 3h38min \\
    \end{tabular}
    \caption{Bulk speed performance metrics for the time range from May 19, 2012 to May 26, 2017. $^1$From CAMEL Scoreboard. Best value for each metric is given in bold.}
    \label{tab:point-to-point-metrics}
\end{table}

\begin{figure}
   \centering
   \subfigure{{\includegraphics[width=\columnwidth]{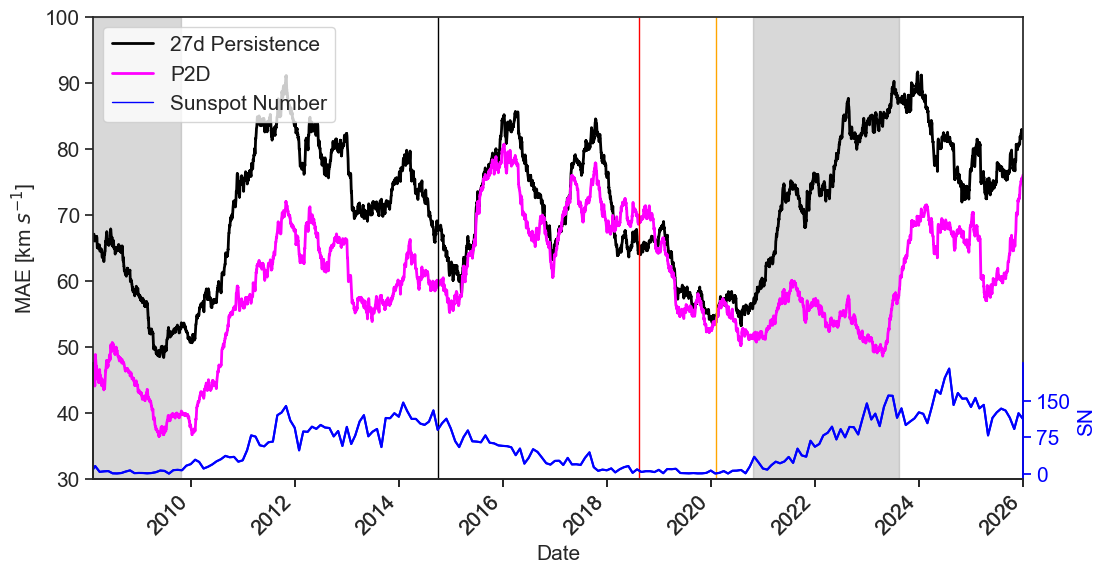}}}

   \caption{\label{fig:MAE_TS_V}\small Running MAE of the speed over the entire time range (2008-2026). Black profile: Running MAE of the 27-day persistence model. Magenta profile: MAE of P2D. Blue profile: Sunspot number. Black vertical line: Loss of contact with STEREO-B. Red vertical line: Launch of PSP. Yellow vertical line: Launch of SolO. The grey shaded areas mark the time periods where either STEREO-B or STEREO-A are located between Earth and L5.} 
   \end{figure}

Figure \ref{fig:MAE_TS_V} shows the performance  evolution of P2D and of the 27-day persistence baseline model over two solar cycles. As we don't compare this analysis with other models, we use here the full time range between October 15, 2008 and January 1, 2026. The magenta line represents the MAE of the bulk speed reconstruction of the P2D model, while the black one shows the MAE of the 27-day persistence model. The lines have been calculated by shifting a 100d window over the model and in-situ profiles and evaluating the MAE for each window. The blue line on the bottom shows the monthly averaged sunspot number (Source: WDC-SILSO, Royal Observatory of Belgium, Brussels, DOI: https://doi.org/10.24414/qnza-ac80) and is scaled to the right axis.

Considering the 27-day persistence reconstruction, we see a trend with the solar cycle. During times of solar maximum, the reconstruction is consistently performing worse than at solar minimum by about 20--30 km~s$^{-1}$ of MAE. There is a strong statistical correlation between the MAE and the sunspot number with a Pearson correlation coefficient of 0.71. For the performance of P2D, we see that it mostly outperforms the persistence reconstruction except for 2015--2020, i.e., the declining phase and solar minimum of Solar Cycle 24, where both models give similar errors. There is a stark difference especially at the solar maxima (2014, 2024), where the performance improves due to the inclusion of further spacecraft and their more advantageous position. It is, however, difficult to conclude a solar cycle dependence for the P2D reconstruction, since it is very dependent on data availability and spacecraft position. The times where it performs best largely coincide with the times where either STEREO-A or STEREO-B are located at the advantageous position between Earth and L5 (grey shaded areas). 

This becomes particularly clear by looking at the period 2023--2026. The  large dip in MAE between 2023 and 2024, is caused mainly by the advantageous position of STEREO-A, which, in this time period, was positioned slightly behind, i.e., east of Earth, resulting in short propagation times from STEREO-A to Earth. This is evident by looking at the large relative contribution from STEREO-A during this period, as is visible in Figures \ref{fig:TS} and \ref{fig:TS_N} and their respective videos. It is the main contributor to the model and its contribution steadily increases as it moves closer to Earth, which is why we can attribute the improved performance mainly to its position. Isolating STEREO-A's contribution we find that its contribution improves the model performance by 14~$km~s^{-1}$ in terms of MAE during this time period. As soon as it overtook Earth in August 2023 and went ahead, i.e., west of Earth, the required propagation time in our model sharply increased to about a full solar rotation, and correspondingly also the MAE rises sharply. When only evaluating this time window, we would also not find a strong correlation between P2D performance and sunspot number, with the Pearson correlation coefficient being only 0.29. 

In addition, in 2018 there is a period where the simple 27-day persistence model slightly outperforms P2D. The reasons for this are unadvantageous constellations of STEREO-A and later PSP. During those times, they were separated far enough from Earth to measure an in-situ profile, which happened to be slightly more dissimilar to future OMNI measurements than the 27 days-old OMNI profile itself. This is not due to a larger latitudinal separation either, rather due to a quick evolution of the solar wind profile during this period, i.e., the assumption of persistence did not hold very well. Therefore, the inclusion of STEREO-A and PSP data caused an increase of the MAE of about 1--3 km~s$^{-1}$. With adequate coverage of the inner heliosphere by multiple spacecraft, this effect is consistently negligible.

\subsubsection{Binary Metrics}

Table \Ref{tab:binary_metric_results} shows the results of the binary metric analysis of high-speed enhancements. Note that ICMEs have not been removed from the datasets for comparability with CAMEL data. The time range here is also May 19, 2012 to May 26, 2017. For this analysis we chose an arbitrary threshold of 450 km~s$^{-1}$ for the bulk speed, above which we consider the solar wind speed to be enhanced. During this time, there were 63 ICMEs \citep{RichardsonCane2024} and 144 CIRs \citep{Koller2025} with solar wind speeds above the threshold. As described in Section \ref{sec:longitudinal}, the ICMEs are also being propagated across longitude in our model, which leads to errors, as ICMEs don't corotate in reality. The majority of high-speed enhancements during the investigated period are however caused by CIRs.

\begin{table}[]
    \centering
    \begin{tabular}{c|c|c|c|c|c|c|c}
         Metric & P2D & 27d Persistence & ESWF & HUXt & WSA 2.2$^1$ &  AmbSoWi-ML$^1$ \\

         TPR & 0.64 & 0.61  & 0.57 & 0.62 & 0.57  & 0.50 \\
         \hline
         FPR & 0.25 & 0.28 & 0.27 & 0.27 & 0.33 &  0.15 \\
         \hline
         TS & 0.48 & 0.44 & 0.45 & 0.46 & 0.33  & 0.38 \\
         \hline
         TSS & 0.39 & 0.32  & 0.30 & 0.36 & 0.24  & 0.35 \\
         \hline
         Bias & 1.00 & 1.00  & 0.84 & 0.97 & 1.27  & 0.82 \\
        
    \end{tabular}
    \caption{Binary bulk speed performance metrics. $^1$From CAMEL Scoreboard. Best value for each metric is given in bold.}
    \label{tab:binary_metric_results}
\end{table}

In the binary metrics, the P2D exhibits the best performance in almost all of the metrics. It beats all the other models in our analysis in terms of correct predictions, with a true positive rate of 0.64. It yields a threat score of 0.48 (combining correct predictions, false alarms, and missed events) and a true skill score of 0.39 (the difference between the true positive and false positive rates). It is tied with the 27-day persistence in terms of bias, where both have a value of 1.00, which means that they each have an equal number of false negatives and false positives. AmbSoWi-ML has the lowest false positive rate with only 0.15. The absolute number depends heavily on the chosen threshold, but the relative performance of the models in comparison to each other is very stable.

\subsubsection{Peak Analysis - CIRs}
\label{sec:CIR}

In addition to high-speed enhancements, which include ICMEs, we evaluate the performance for actual CIRs. To this end, we use the high-speed stream catalog provided by \cite{Koller2025}. In addition to the speed peaks, the high-speed streams produce CIRs with proton density peaks, both of which we attempt to reconstruct. Here, we only compare to models, which we have data access to and can calculate the metrics ourselves due to our strict constraints of only considering CIRs, leaving us with P2D, the 27-day persistence model, ESWF and HUXt. Due to these restrictions, the total number of peaks we investigate is 330 between May 2012 and January 2024. Figures \ref{fig:Peaks_1} to \ref{fig:Peaks_4} in the appendix show the entire bulk speed time series along with the CIR times. 

The accuracy of peak prediction during CIRs is shown in Table \ref{tab:CIR_results}. A speed peak is considered as a hit, as long as the modeled peak is within $\pm 100~km~s^{-1}$ in amplitude and $\pm 1~$day in timing of the measured peak.

\begin{table}[]
    \centering
    \begin{tabular}{c|c|c|c|c|}
         Metric & P2D & 27d Persistence & ESWF & HUXt \\
         \hline
         ME - timing & 0.29 d & 0.17 d & $-$0.03 d & 0.25 d \\ 
         \hline
         RMSE - timing & 1.20 d & 1.21 d & 1.78 d & 1.41 d \\
         \hline
         ME - speed & $-$29.90 km~s$^{-1}$ & $-$10.28 km~s$^{-1}$ & $-$2.49 km~s$^{-1}$ & $-$14.67 km~s$^{-1}$\\ 
         \hline
         RMSE - speed & 78.66 km~s$^{-1}$ & 93.21 & 103.82 km~s$^{-1}$ & 84.57 d \\ 
         \hline
         Hit Rate & 0.51 & 0.51 & 0.37 & 0.42\\

    \end{tabular}
    \caption{CIR bulk speed performance metrics. Best value for each metric is given in bold.}
    \label{tab:CIR_results}
\end{table}

P2D performs well in comparison to the other models in terms of RMSE and hit rate, defined as the amount of hits divided by the total amount of peaks. There, it is tied for first place along with the 27-day persistence model. P2D has an RMSE of 1.2 days in timing of the peak and 78.66 km~s$^{-1}$ in amplitude, both of which are lowest values of the four models. ESWF performs best in terms of average peak timing, with a value of $-$0.03 days as well as average peak amplitudes with a value of $-$2.49 km~s$^{-1}$.

\begin{figure}
   \centering
   \subfigure{{\includegraphics[width=\columnwidth]{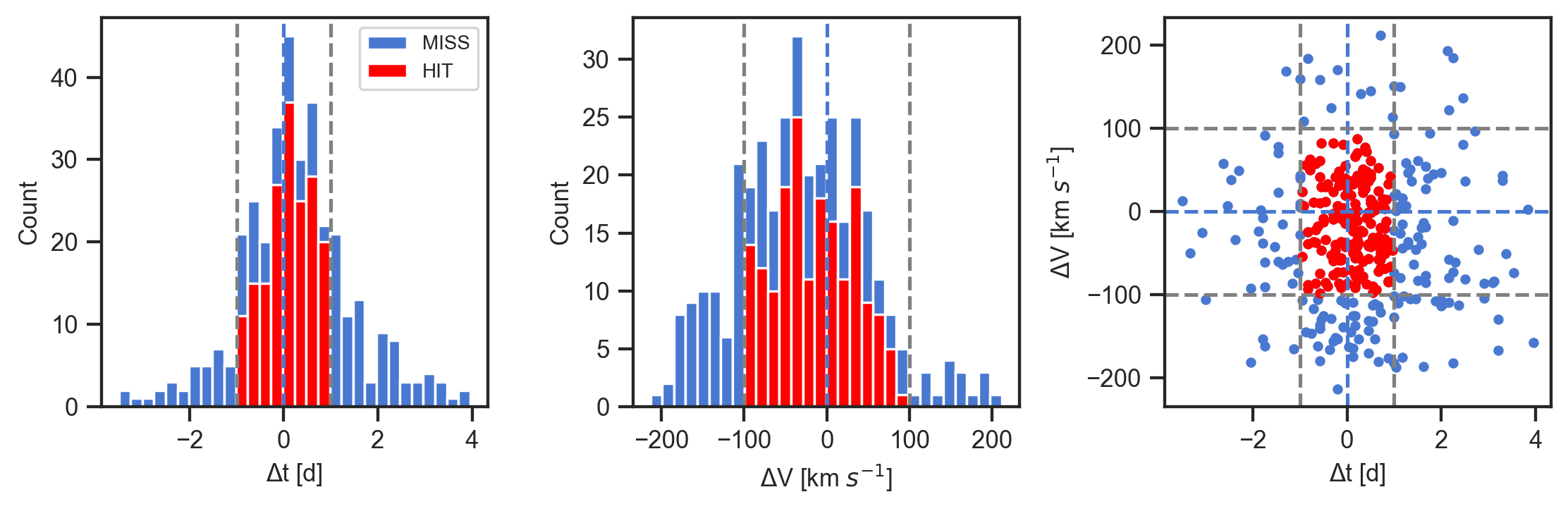}}}

   \caption{\label{fig:Peaks_stats}\small Speed peak error distributions. Left: Histogram of timing errors. Center: Histogram of speed errors. Right: Scatterplot of speed errors against timing errors. Blue: Misses. Red: Hits} 
\end{figure}

Figure \ref{fig:Peaks_stats} shows the distributions of bulk speed and timing errors of the CIR peak speeds. The first panel shows the distribution of peak timing errors, with a slightly larger flank on the right side, which agrees with the positive average timing error of 0.29 days. The central panel shows the distribution of the amplitude errors, with a visible skew to lower velocities. This means that peak speed is often underestimated, which also leads to a large average speed error of $-$29.9 km~s$^{-1}$. Both of these effects are noticeable in the right panel, where the distribution is slightly skewed to the lower right, i.e., lower speeds and later times. 

\subsection{Results - Proton Density}

The data availability for proton density models is much more scarce than for bulk speed. We compare point-to-point metrics and density peak performance only to models where we have access to an entire time series. This leaves us with only P2D and the 27-day persistence model between January 2008 and January 2024.  

\subsubsection{Point-to-point Metrics}

Table \ref{tab:point-to-point-metrics_N} shows the performance of the proton density model of P2D compared to a 27-day persistence model.

\begin{table}[]
    \centering
    \begin{tabular}{c|c|c|}
         Metric & P2D & 27-day persistence \\
         \hline
         ME& $-$0.67 & $-$0.08    \\
         \hline
         MSE& 30.89 & 37.30   \\
         \hline
         MAE& 3.51  &  3.89   \\
         \hline
         RMSE& 5.56 & 6.11  \\
         \hline
         CC& 0.21 & 0.18   \\
         \hline
         Cadence & 1h & 1h  \\
    \end{tabular}
    \caption{Density performance metrics. Best value for each metric is given in bold.}
    \label{tab:point-to-point-metrics_N}
\end{table}

Except for mean error, where P2D underestimates the mean density by 0.67 cm$^{-3}$ and the 27-day persistence model only by 0.08 cm$^{-3}$, P2D performs better in all other metrics. P2D has an MAE of 3.51 cm$^{-3}$ and an RMSE of 5.56 cm$^{-3}$, both of which are slightly better than the 27-day persistence model. For context, the in-situ measured average density during the entire time range is 6.25 cm$^{-3}$ and its standard deviation is 4.79 cm$^{-3}$. Therefore, the RMSE is almost as large as the average density value, indicating a bad performance. This is also evident by looking at the values of the Pearson correlation coefficients between predicted and measured density values, being  0.21 for P2D and 0.18 for the 27-day persistence model, respectively. 

   \begin{figure}
   \centering
   \subfigure{{\includegraphics[width=\columnwidth]{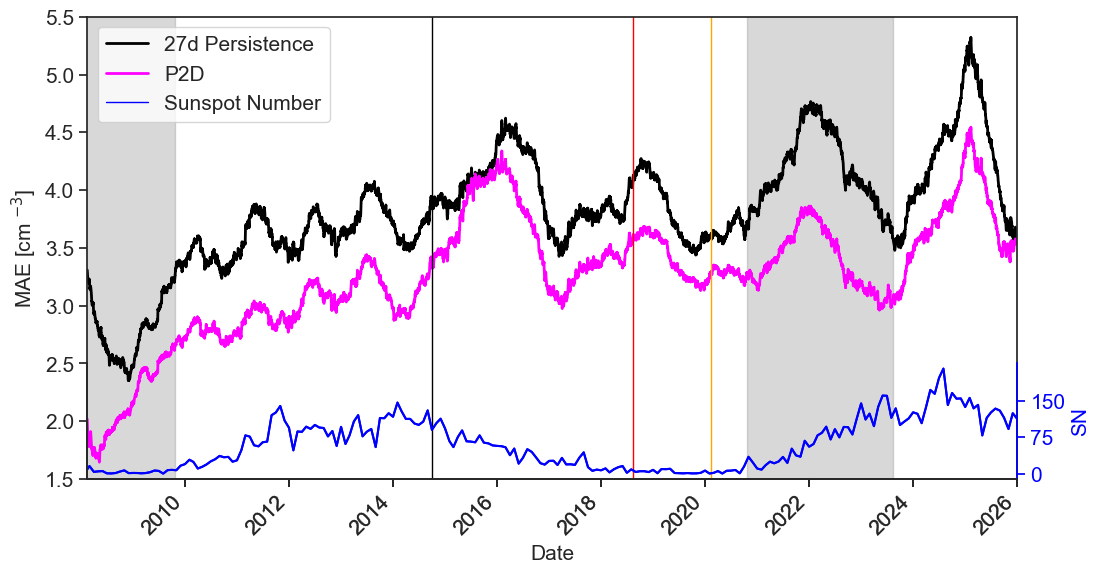}}}

   \caption{\label{fig:MAE_TS_N}\small Running MAE of the proton density over the entire time range (2008-2026). Black profile: Running MAE of the 27-day persistence model, Magenta profile: MAE of P2D, Blue profile: Sunspot number. Black vertical line: Loss of contact with STEREO-B, Red vertical line: Launch of PSP, Yellow vertical line: Launch of SolO. The grey shaded areas mark the time periods where either STEREO-A or STEREO-B are located between Earth and L5.} 
   \end{figure}

Figure \ref{fig:MAE_TS_N} shows the MAE of the density model as a time series over the entire time range, similar to Figure \ref{fig:MAE_TS_V}. We see consistently that P2D outperforms the 27-day persistence model over both solar cycles. There is no clear trend with solar activity (see Section \ref{sec:Discussion}).  

\subsubsection{Peak Analysis - CIRs}
\label{sec:CIR_N}

Figures \ref{fig:Peaks_N1} to \ref{fig:Peaks_N4} in the appendix show the entire density time series along with the CIR times. 

Table \ref{tab:CIR_results_N} shows the results of the density peak analysis of P2D compared to the 27-day persistence model. A density peak is considered as a hit, as long as the modeled peak is within $\pm 10$ cm$^{-3}$ in amplitude and $\pm 1$~day in timing of the measured peak.

\begin{table}[]
    \centering
    \begin{tabular}{c|c|c}
         Metric & P2D & 27-day persistence \\
         \hline
         ME - timing & 0.03 d & 0.01 d  \\ 
         \hline
         RMSE - timing & 0.74 d & 0.83 d  \\ 
         \hline
         ME - density & $-$6.53 $cm^{-3}$  & $-$5.38 $cm^{-3}$ \\ 
         \hline
         RMSE - density & 13.84 $cm^{-3}$& 14.48 $cm^{-3}$  \\ 
         \hline
         Hit Rate & 0.48 & 0.45\\
        
    \end{tabular}
    \caption{CIR density performance metrics. Best value for each metric is given in bold.}
    \label{tab:CIR_results_N}
\end{table}

Interestingly, the timing performances of the density peaks are great, with an average timing error of 0.03 days and 0.01 days for P2D and the 27-day persistence model, respectively. The peak density, however, is being systematically underestimated, with average errors of $-$6.53 $cm^{-3}$ for P2D and $-$5.38 cm$^{-3}$ for the 27-day persistence model. The hit rates are 0.48 and 0.45, respectively. 

\begin{figure}
   \centering
   \subfigure{{\includegraphics[width=\columnwidth]{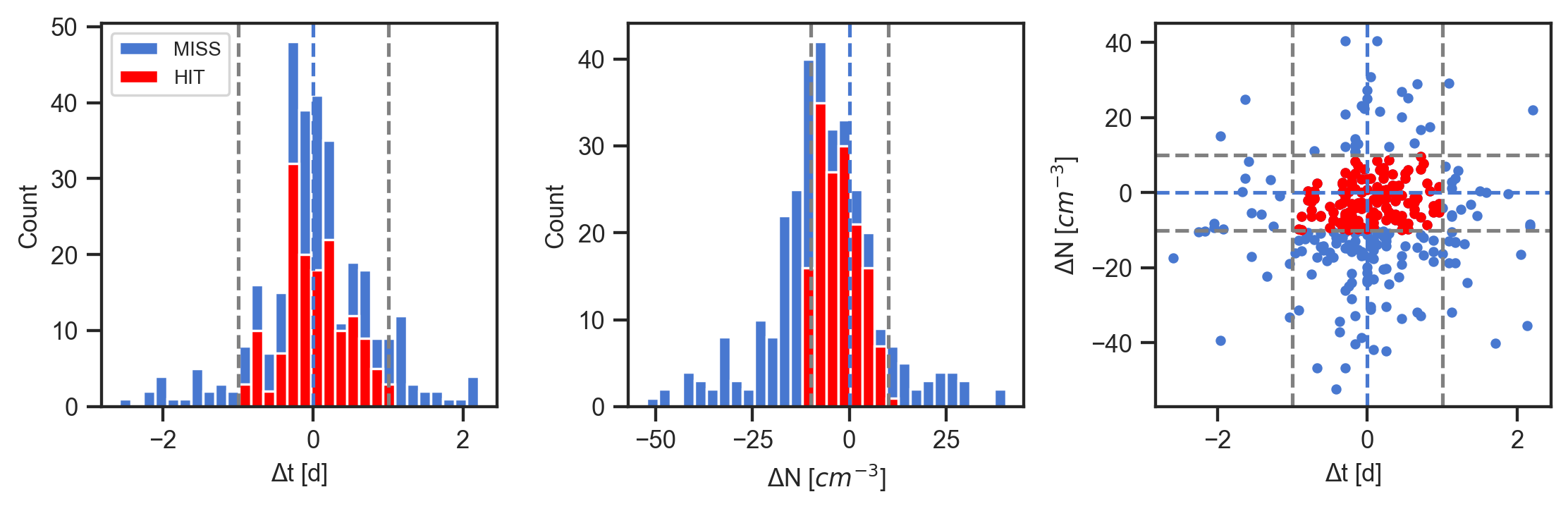}}}

   \caption{\label{fig:Peaks_Nstats}\small Density peak error distributions. Left: Histogram of timing errors. Center: Histogram of density errors. Right: Scatterplot of density errors against timing errors. Blue: Misses. Red: Hits} 
   \end{figure}

These errors are again visualized in Figure \ref{fig:Peaks_Nstats}. The left panel shows good average peak timings with a small standard deviation (RMSE) of 0.73 days. The central panel is skewed to the left, meaning underestimated peak amplitudes. This is also clearly visible in the right panel, where many blue dots, i.e. misses, are below the $-$10 cm$^{-3}$ line, which we chose as a threshold.  

\section{Discussion}
\label{sec:Discussion}

\subsection{Lead time}
As our model is based on persisting solar wind, which is assumed to circulate the Sun infinitely, we can choose the lead time on the model arbitrarily. Throughout this work, we present a reconstruction with zero lead time, i.e., the reconstruction has access to all science-level data as input to date at the reconstructed time.
Of course, as the lead time increases, the uncertainty and unreliability of the model will also increase, which renders lead times of multiple months useless (see Panel (b) of Figure \ref{fig:N_COR} as well as \cite{Milosic2026}).

This minimizes our dependence on timely data availability. In an operational setting we would have to use near-real time data, which excludes PSP due to its large data-latency. For other spacecraft, gaps in their data could be filled by slightly older data from other spacecraft. 

The influence of lead time on the performance of P2D can be estimated from Figure \ref{fig:MAE_TS_V}. The magenta line is the performance of P2D, assuming real-time data availability. If we incrementally increase the lead time, fewer spacecraft would be available to contribute. This way, the purple line would morph step-by-step into the black line as we approach a lead time of 27 days.

\subsection{Applications}

There are two types of P2D model output: (1) The one dimensional output at one specific location and (2) the two dimensional maps of solar wind data projected on to the ecliptic. 

The one-dimensional output can be used as a solar wind forecast at Earth or any other position of interest in the solar system, as long as there is a spacecraft at closer distances to the Sun. \cite{Milosic2026} showed that analogue error profiles can be produced at the position of Mars. Therefore, a solar wind forecast could also be implemented at other planets, conceivably up until Jupiter and beyond. Depending on spacecraft measurement availability, this would however lead to larger errors if the data is being propagated for greater distances.

There are also multiple ways to use the two-dimensional model output of P2D. One such way would be the scientific investigation of CME propagation through the ambient solar wind, similar to \cite{Baratashvili2025, Zhang2026}. As the 2D output gives us snapshots of the ambient solar wind conditions, it is possible to investigate the impact of the solar wind on the trajectory of the CME, although latitudinal information about the solar wind would be very limited. 

Another way to use the 2D output is in combination with other models. Starting from the P2D output based on multiple solar probes, one could run HD and MHD models with real in-situ measurements as boundary conditions as has been done by \cite{Owens2026}. This has the advantage of not relying on magnetograms and coronal models as input. The 2D output could also be used in synergy with Interplanetary scintillation-based tomography \citep{Tiburzi2023}, which can reconstruct 3D information on the solar wind.

The great advantage of P2D lies in combining speed and density results. This allows for the calculation of dynamic pressure, a key parameter for space weather forecasting alongside $B_z$ \citep{Gonzalez1989, Veenadhari2025}. Furthermore, the resulting 2D density map enables more precise, distance-dependent drag force calculations. This aligns with MHD models like EUHFORIA and ENLIL, which utilize density enhancement factors (\textit{dcld}) -- a parameter shown to significantly improve CME propagation forecasts \citep[][]{Werner2019, Yordanova2024}.

\subsection{Limitations}

P2D applies a very simplified approach to solar wind modeling which comes with a few limitations. (1) Radial evolution is limited to a simple inelastic collision scheme, which doesn't reflect the behavior of solar wind plasma accurately. \cite{Milosic2025} showed that this scheme can achieve reasonable reconstructions, but we expect the errors to increase with applications to further radial distances. (2) P2D has no explicit treatment of ICMEs. In this study, ICMES were not removed to make the reconstruction comparable to as many other models as possible. They are therefore corotated just like the ambient solar wind and not treated separately. (3) Although P2D doesn't rely on magnetograms, it is very dependent on the position of other spacecraft. The solar wind can only be reconstructed in regions where there are in-situ measuring spacecraft closer to the Sun and reasonably near the solar equator. (4) The empirical error estimation was performed on a dataset with a maximum latitudinal difference of roughly 15$^{\circ}$ between spacecraft. Therefore, the validity of the error relations used to weigh the individual spacecraft contributions doesn't extend to higher latitudes.

\subsection{Operational Setting}

If P2D were to be employed in an operational setting, two main modifications would need to be made. (1) The analysis in this work was performed on science-level data, which is not available near real-time and therefore cannot be used in an operational setting. At L1, there are multiple sources of near real-time data, like Deep Space Climate Observatory \citep[DSCOVR,][]{DSCOVR}, ACE \citep{ACE}, and Space Weather Observations at L1 to Advance Readiness - 1 (SOLAR-1). Even with multiple days of delay, they could be used as input, since their data would only be used for the next Carrington rotation, i.e. a lead time of 27 days. STEREO-A beacon data is also available near real-time and could be used as input. Similarly, SolO low-latency data could be used as input with a slight delay. As long as it is not located close to the Sun-Earth line, it would also provide valuable input to the forecast. PSP data is usually not available for months after measurement and could therefore not be used for the real-time forecast. As soon as Vigil \citep{West2025} becomes operational it will provide a good baseline of in-situ measurements to the forecast even if they were to be available with up to three days of delay. (2) ICMEs were not treated separately in this work. For a reasonable forecast, it would be good to remove data during ICME encounters, as they do not corotate, which is an assumption of P2D. ICMEs would have to be automatically detected as they are encountered by the spacecraft \citep[see, e.g.,][]{Rudisser2026} and removed. The gaps that they leave could be filled up with spacecraft measurements at different positions, which did not encounter an ICME.

\subsection{Future prospects}

The performance of P2D is strongly dependent on spacecraft measurements. The positions and data availability of spacecraft can drastically impact the model's metrics, as is evident from Figure \ref{fig:MAE_TS_V}. Because of this, the MAE can vary between 40 km~s$^{-1}$ and 80 km~s$^{-1}$.

SolO is slowly moving out of ecliptic, increasing its latitudinal separation to Earth for large parts of its orbit. This will cause a reduction of SolO's contribution to the P2D forecast according to Equation \ref{eq:MAE_V} and \ref{eq:MAE_N}, making SolO less useful for the model. 

Vigil is a space mission planned to be stationed at the L5 point \citep{West2025}. This position is strategically advantageous, as it is 'upstream' in the sense of solar rotation. If the mission's orbit is carefully planned to avoid high latitudinal separations from Earth, this could be a strong baseline for the P2D forecast. A stable position upstream of Earth would provide us with a good estimate of the solar wind with four to five days of lead-time. The grey-shaded areas in Figure \ref{fig:MAE_TS_V} show times, where either STEREO-A or STEREO-B are located upstream between the Earth and L5. During these times the performance of P2D is best with MAEs between 40 km~s$^{-1}$ and 60 km~s$^{-1}$. Although positions between L5 and Earth are even more advantageous, L5 also shows a good performance and there is no stable parking position for spacecraft in between.

In addition to L5, there have been proposals for space missions between L1 and the Sun, so called sub-L1 missions \citep{SWIFT2023, Lugaz2025}, which would greatly improve P2D's performance as well. The quasi-zero longitudinal separation from Earth is expected to produce an MAE in speed of roughly 40 km~s$^{-1}$ \citep[see discussion in][]{Milosic2026} which would be unprecedented in solar wind forecasts. Lead times, however, would be very short, i.e., in the range of a few hours.

\subsection{Model comparison}

P2D, with a bulk speed MAE of 65.41km~s$^{-1}$, consistently outperforms the 27-day persistence reconstruction in almost all metrics. Furthermore, it outperforms all other models in terms of binary bulk speed metrics (see Table \ref{tab:binary_metric_results}). It performs best when there is a spacecraft (STEREO-A or B) located between Lagrange point L5 and Earth. During these times the bulk speed MAE decreases by roughly 35\% as compared to the 27-day persistence model. 

The only model which consistently performs better than P2D is the machine learning model AmbSoWi-ML \citep{Bailey2021}. AmbSoWi-ML is trained on coronal models, using the WSA approach \citep{Arge2000}, as well as the OMNI data set. It is therefore ultimately based on magnetograms as input, like WSA, with a lead time of four days. Nevertheless, 
AmbSoWi-ML performs best in terms of point-to-point metrics compared to the other models.

During CIRs (Tables \ref{tab:CIR_results} and \ref{tab:CIR_results_N}), P2D systematically underestimates peak speeds and densities due to its profile-averaging nature (see Eqations. \ref{eq:Contributions} and \ref{eq:Combination}). ESWF \citep{ESWF2007, ESWF2015, ESWF2016, ESWF2023} uses coronal hole areas as input calculates the solar wind speed and timing at Earth based on an empirical relationship. Being fine-tuned to correct CIR speed predictions, it performs best at speed peaks during these events. 

In general, all of the available density reconstructions perform badly. The MAE in density is 3.51cm$^{-3}$, which is as large as 50\% of the mean density during the investigated period. Interestingly, the timing of the peak density performs better than the timing of the peak speed. Ahead of high speed streams, density often exhibits a sharp, clear 'compression' feature \citep[see, e.g.,][]{Richardson2018}. This sharp rise creates a well-defined time stamp for density that is relatively easy to locate. On the other hand, high-speed streams speed profiles are often governed by complex, large-scale, and long-duration expansion processes, which makes it harder to find precise maxima. 

\subsection{Solar Cycles and Performance}

While we find a clear trend with the solar cycle for the 27-day persistence forecast ($cc=0.71$), there is no strong trend for P2D ($cc=0.29$). During solar minimum, the large-scale structure of the solar wind is less dynamic with fewer CMEs and longer persisting CIRs \citep{Webb1991, Richardson2018}. Therefore, we would expect P2D to show a better performance during those times. We do, however, have consistently better spacecraft coverage during the solar maxima. During the maximum of solar cycle 24 (around 2014), we had three spacecraft at roughly 1AU at different longitudes. During the maximum of solar cycle 25 (around 2024), we had three spacecraft at roughly 1AU at different longitudes, while there were only the OMNI database and STEREO-A between those maxima. The increased performance of P2D with multiple spacecraft partly made up for the decreased performance during maxima, reducing the anti-correlation between the sunspot number and the performance of P2D.

P2D performs better than the 27-day persistence forecast especially during solar maxima due to a combination of effects. First, as was stated before, during the maxima there happens to be a greater abundance of spacecraft. Second, there are typically more ICMEs during solar maxima, which would decrease the performance of both persistence models. However, because P2D performs an averaging over multiple spacecraft profiles, the occasional encounter of an ICME of an individual spacecraft is smoothed out by the other spacecraft's profiles.

\subsection{Boundary conditions}

The strong performance of the persistence models in comparison with more sophisticated ones underlines the importance of accurate boundary conditions. P2D, as well as the 27-day persistence model both use in-situ measurements from spacecraft as input in order to model in-situ measurements at future times. This direct correspondence between model input and output minimizes the amount of physical assumptions necessary, leading to a better performance. It's greatest advantage is the reliability of the boundary conditions, whereas magnetograms, which are used as boundary conditions for many other models, are subject to a number of problems \citep[See Section 4.1 in][]{MacNeice2018}. 

\subsection{Dependence on latitude and longitude}

Panel (a) of Figure \ref{fig:N_COR} shows the dependence of the MAE of the proton density as a function of latitudinal spacecraft separation. The relationship appears to be linear up until a separation of roughly 5$^\circ$, after which the density error doesn't increase. These findings agree with \cite{Owens2020} and \cite{Turner2021}, who also find such a threshold of 5$^\circ$. \citep{Chakraborty2023} find a causal relationship between both the latitude and lead-time on the MAE in solar wind speed, with more emphasis on the lead-time. For speed, \cite{Milosic2026} find a linear relationship of the MAE even above 5$^\circ$ in latitudinal separation. In lead-time there is a logarithmic relationship, showing larger effect at low lead-times. 

\subsection{Dependence on radial distance}

As for radial distance, we find no significant relationship with the MAE. We would however expect radial distance to play a role, as there are multiple effects in the radial direction changing the profiles of the solar wind parameters. We attribute the lack of a trend to the clustering of data around 1AU, because of the large amounts of data from the STEREO mission and OMNI, and the comparatively low amounts of data from SolO and PSP.

\section{Conclusions}

We presented a new approach to solar wind persistence modeling, based on in-situ measurements from multiple spacecraft in the inner heliosphere. Spacecraft measurements are being propagated across longitudinal and radial distances, producing two-dimensional maps of solar wind speed and density. Longitudinally, we use the persistence of the solar wind and propagate the data with an angular speed equivalent to the solar rotation rate. Radially, we propagate the data ballistically with inelastic collisions at interaction regions. These maps can be evaluated at any target location, as shown here for Earth, and combined according the errors based on their longitudinal and latitudinal separation to the target location. 

Our approach offers multiple advantages over other solar wind models:

\begin{itemize}
    \item (1) It provides a good overall reconstruction of the bulk speed, with an MAE of only 65.41 km~s$^{-1}$ as well as a true positive rate of 64\% during high-speed enhancements. The mean error of the timing of CIR speed peaks is 0.29~days and their speed maximum is underestimated by 29.9~$km~s^{-1}$ on average. Regarding the proton density reconstruction, it beats the 27-day persistence model with an MAE of 3.51~$cm^{-3}$ and the timing of CIR proton density peaks is very accurate with a mean error of less than an hour. The peak proton density during CIRs is underestimated by 6.53~$cm^{-3}$ (see Tables \ref{tab:point-to-point-metrics}--\ref{tab:CIR_results_N}).
    \item (2) Since the model is based on persistence of in-situ data, we can arbitrarily decide on a lead-time, which, however, will affect the performance.
    \item (3) The model doesn't depend on computationally expensive calculation and can therefore easily be implemented operationally.
    \item (4) In contrast to other persistence models, the simulation output can be evaluated at any point in the heliosphere, as long as there is at least one spacecraft at a smaller distance to the Sun. 
    \item (5) The model offers a density reconstruction, which can be used in combination with bulk speed for calculating the dynamic pressure. 
\end{itemize}

As of yet, P2D is not used operationally, but we aim to make the code and the real-time results publicly available.

\begin{funding}
This research was funded in whole or in part by the Austrian Science Fund (FWF) [10.55776/PAT2280725]. SGH acknowledges funding from the Research Council of Finland (Academy Fellowship) [370747; RIB-Wind]. This research was funded in whole or in part by the Austrian Science Fund (FWF) [10.55776/PAT2280725]. S.J.H. was supported in part by the National Science Foundation grant AGS-2229100.
\end{funding}

\begin{conflictofinterest}
The authors declare no Conflict of Interest.
\end{conflictofinterest}

\begin{dataavailability}
Data will be made public.
\end{dataavailability}


\bibliography{jswsc}

\newpage

\begin{appendix} 

\section{Density Error Estimation}

\begin{figure}[hb]
   \centering
   \subfigure{\includegraphics[width=\columnwidth]{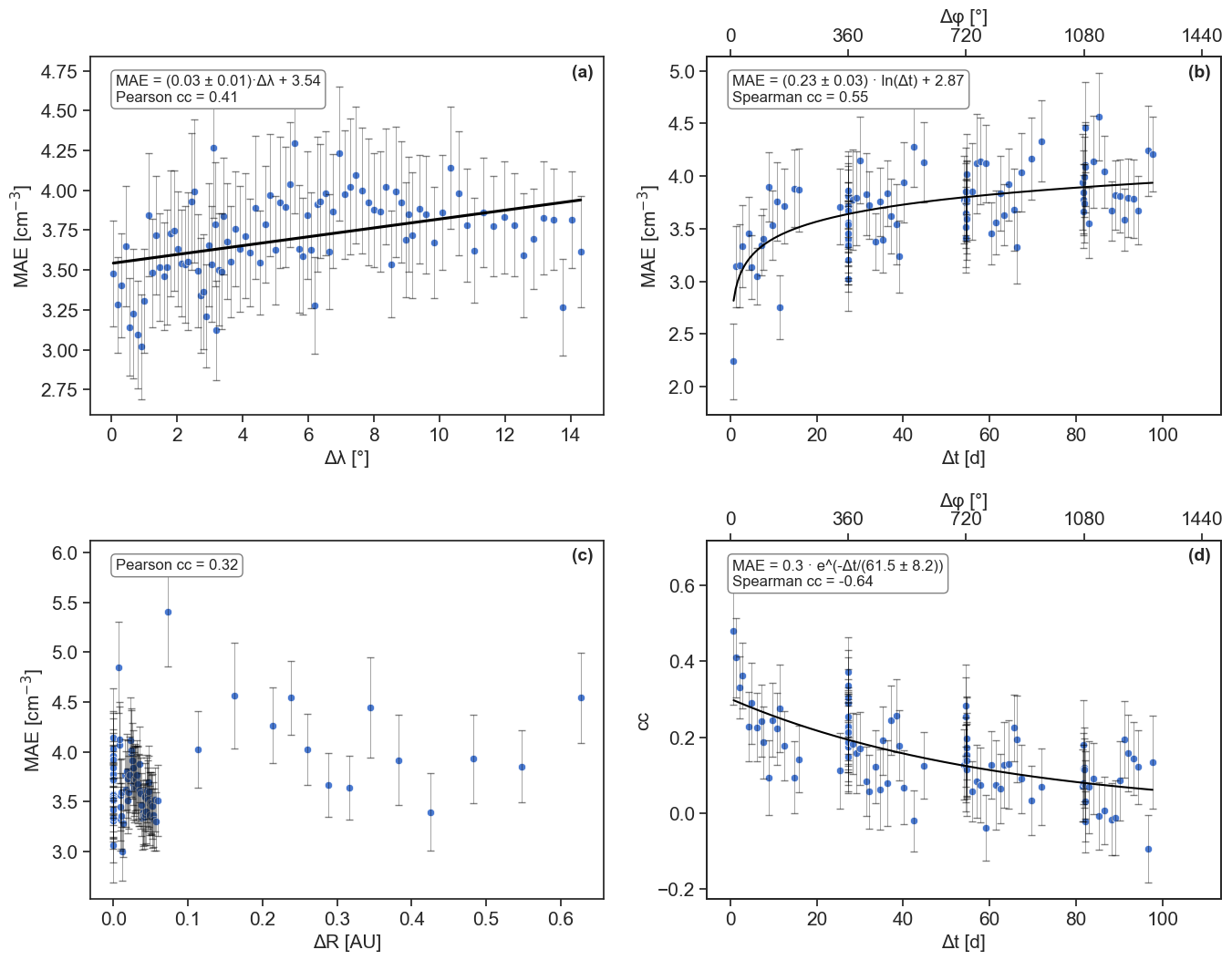}}

   \caption{\label{fig:N_COR}  \small MAE and CC of density in dependence of different dimensions of spacecraft separation. For a detailed description of this analysis as applied to the speed component, we refer to \cite{Milosic2026}.} 
   \end{figure}

Figure \ref{fig:N_COR} shows the Mean Absolute Error (MAE) across (a) latitudinal separation, (b) longitudinal separation and (c) radial separation. This analysis was performed over dataset including OMNI, STEREO-A, PSP and SolO between 2014 and 2024. Their data were propagated across two dimensions in the same scheme as described in Sections \ref{sec:longitudinal} and \ref{sec:radial}. Then, the data were combined into bins of 1000 data points each and plotted against the distance they have traveled across radial distance, longitude and latitude. We find similar trends for the density as for the speed, with linear effect of latitude, logarithmic dependence on longitude, and no trend for radial distance. 

\section{Spacecraft Orbits and Model Composition}

\begin{figure}
   \centering
   \subfigure{{\includegraphics[width=\columnwidth]{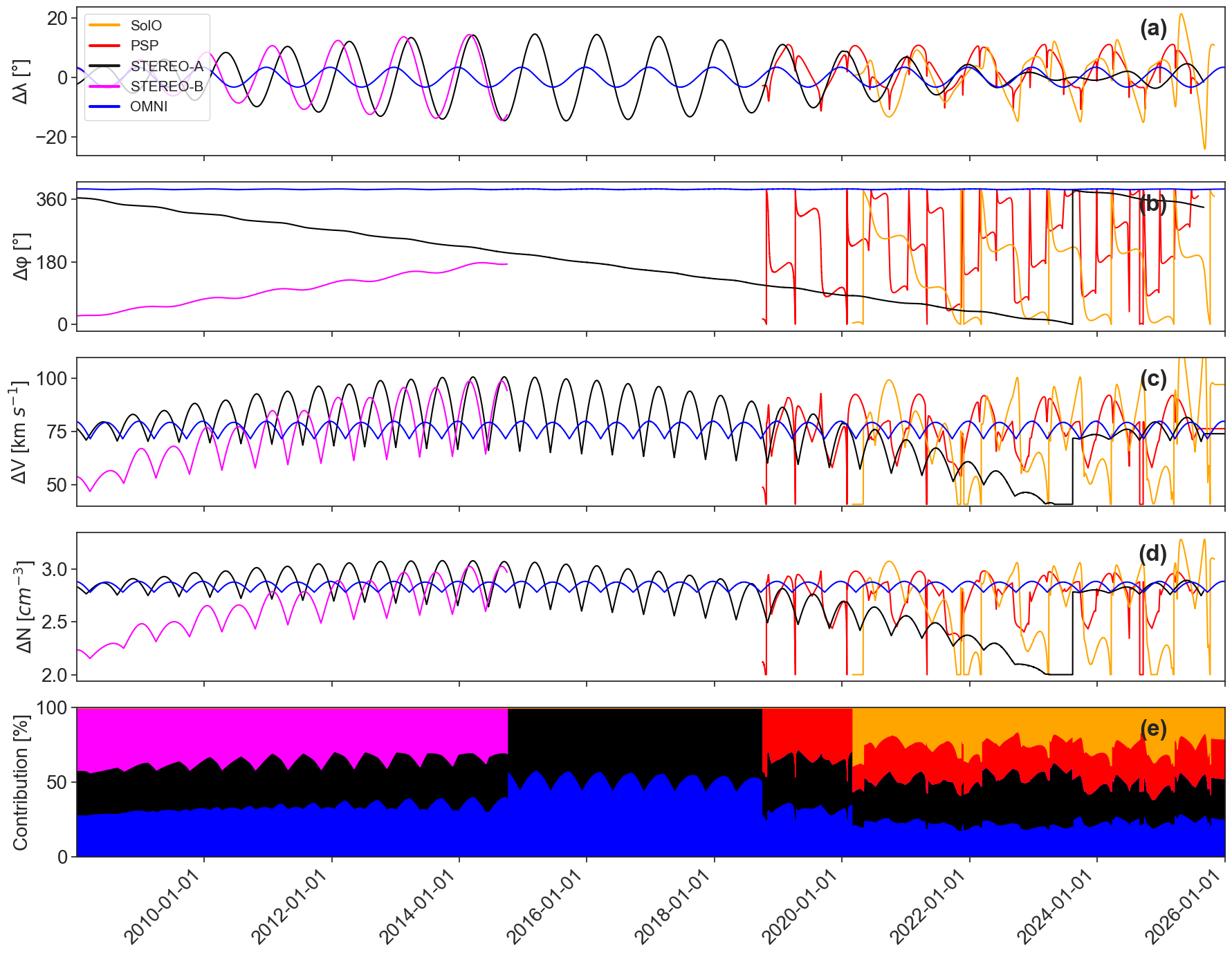}}}

   \caption{\label{fig:Spaghetti_long}\small Spacecraft orbits and resulting reconstruction uncertainties and contributions for the entire time range. Yellow: SolO; Red: PSP; Black: STEREO-A; Magenta: STEREO-B; Blue: OMNI. Panel (a): Latitudinal separations to Earth; (b): Longitudinal separations to Earth; (c): Speed uncertainty; (d): Density uncertainty; (e): Relative contributions of each spacecraft to the reconstruction.}
   \end{figure}

Figure \ref{fig:Spaghetti_long} shows how the spacecraft orbits affect the combination of time series into the near-Earth prediction, exactly like Figure \ref{fig:Spaghetti} in the main body, but for the time range between 2008 and 2026. Panels (a) and (b) show the latitudinal and longitudinal separation of the input spacecraft to Earth. Note that the longitudinal separation exceeds 360$^\circ$ in HEEQ, due to Earth's longitudinal motion during a Carrington rotation. Panels (c) and (d) show the speed and density uncertainties, resulting directly from evaluating the data from Panels (a) and (b) with Equations \ref{eq:MAE_V} and \ref{eq:MAE_N}. The bottom Panel, (e), shows the individual spacecraft contributions to the speed prediction as a percentage. In August 2023, STEREO-A moves from the East of Earth to its West, producing a sudden jump in the uncertainty, as the longitudinal separation increases suddenly from roughly 0$^{\circ}$ to slightly more than 360$^{\circ}$.

\section{Time series}

\begin{figure}
   \centering
   \subfigure{{\includegraphics[width=\columnwidth]{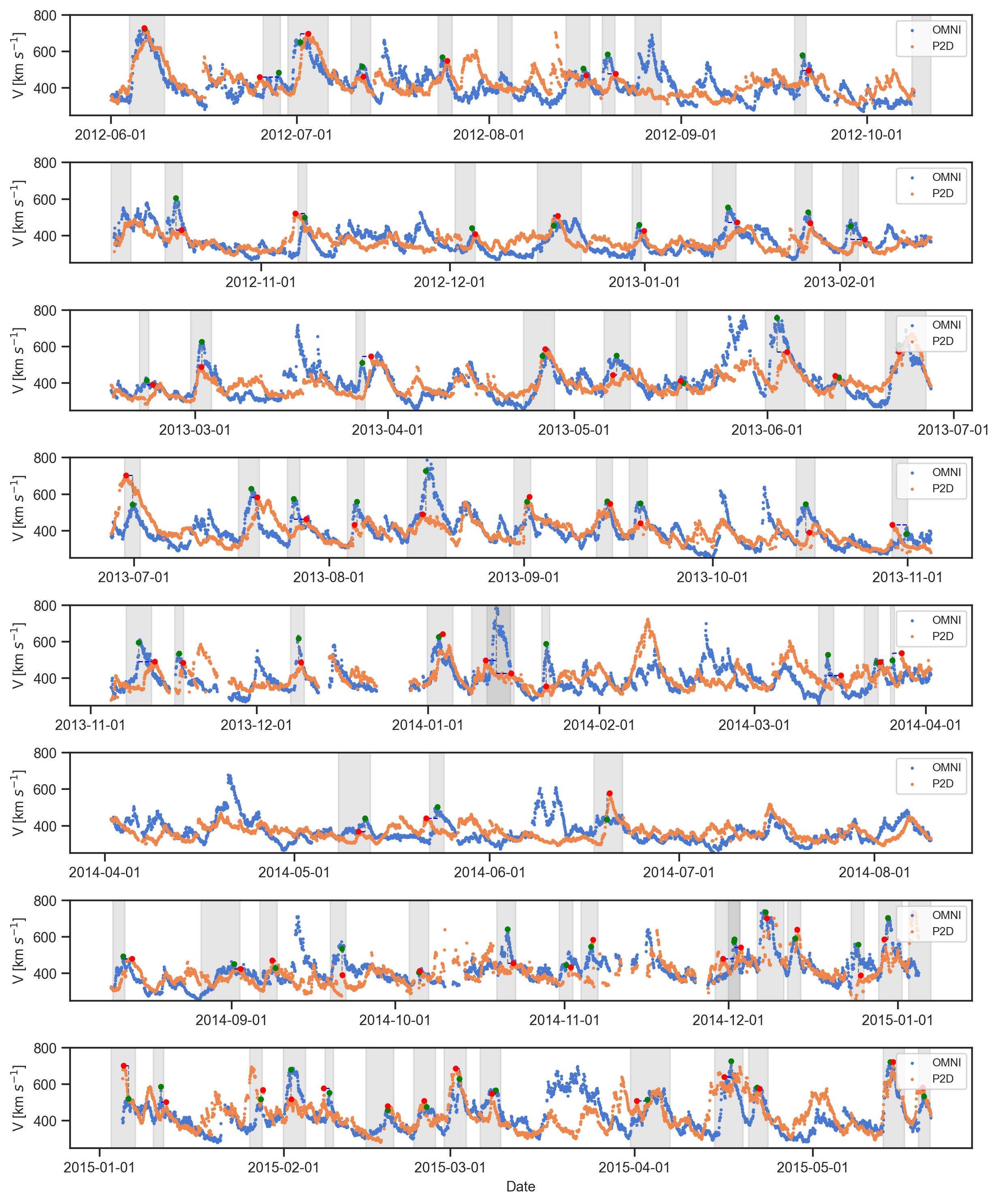}}}

   \caption{\label{fig:Peaks_1}\small Solar wind bulk speed time series with CIR intervals (grey shaded areas) between June 2012 and June 2015. Blue: OMNI in-situ measurements; orange: Model reconstruction.}
   \end{figure}

\begin{figure}
   \centering
   \subfigure{{\includegraphics[width=\columnwidth]{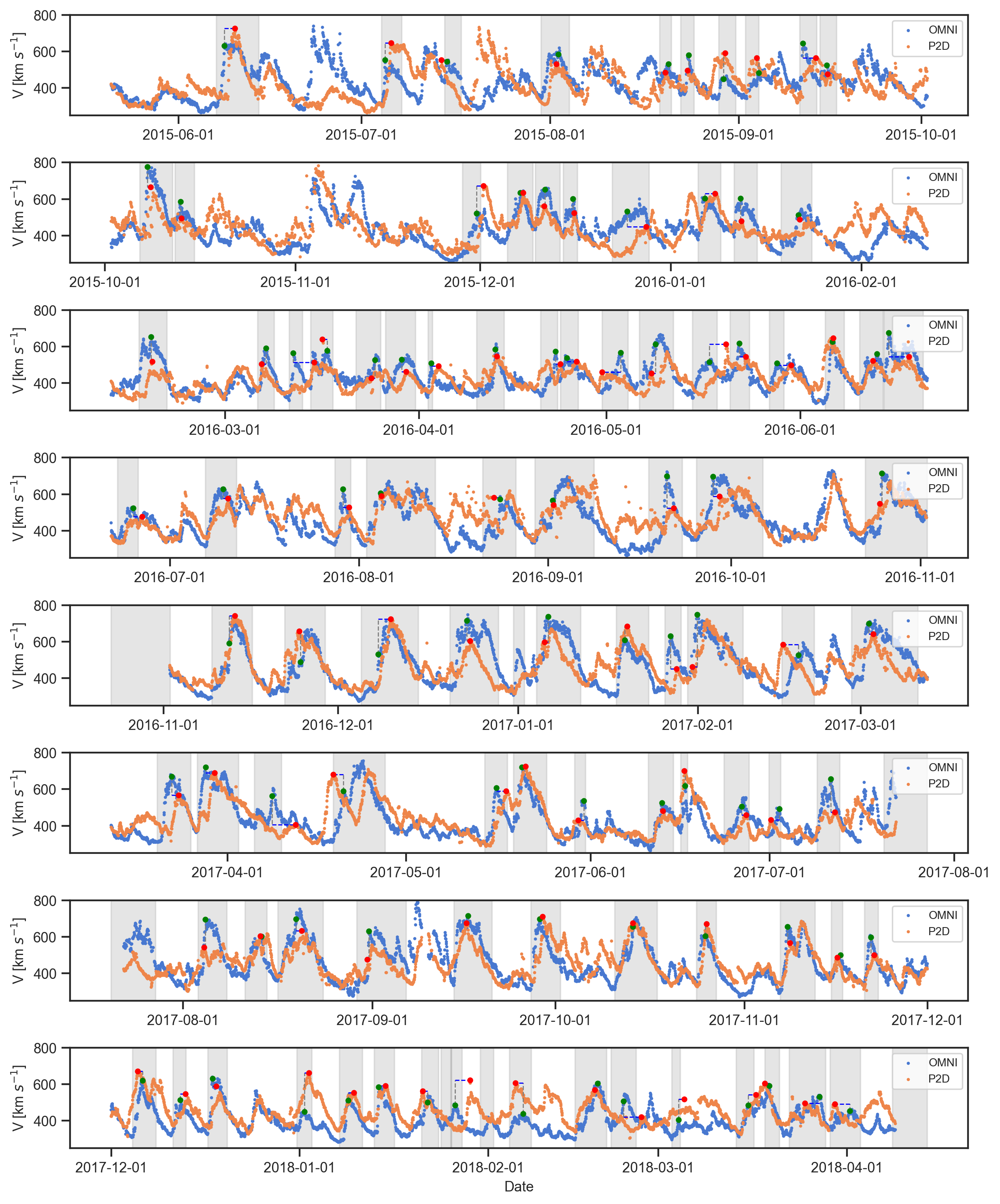}}}

   \caption{\label{fig:Peaks_2}\small Solar wind bulk speed time series with CIR intervals (grey shaded areas) between June 2015 and April 2018. Blue: OMNI in-situ measurements; orange: Model reconstruction.} 
   \end{figure}

\begin{figure}
   \centering
   \subfigure{{\includegraphics[width=\columnwidth]{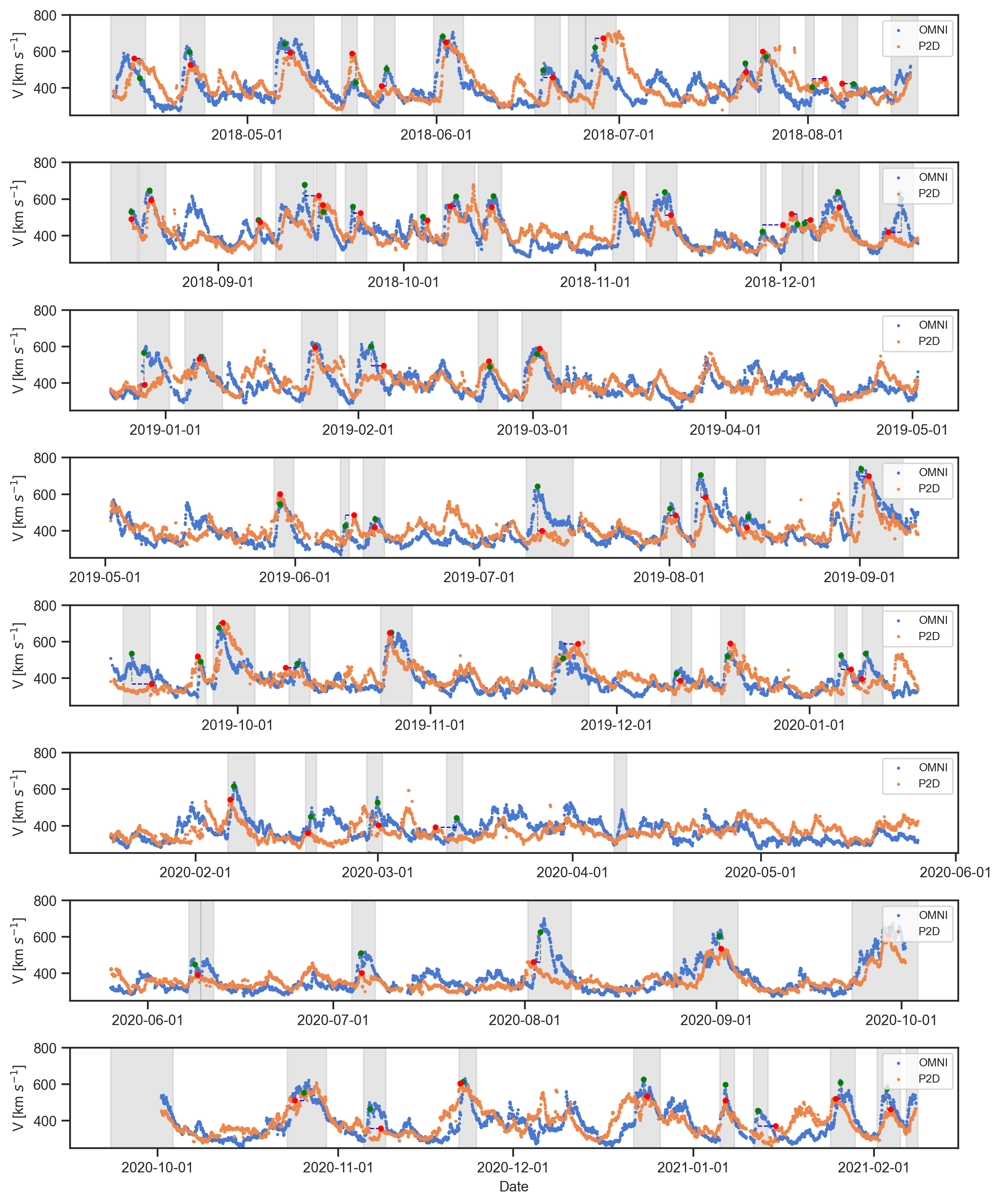}}}

   \caption{\label{fig:Peaks_3}\small Solar wind bulk speed time series with CIR intervals (grey shaded areas) between April 2018 and February 2021. Blue: OMNI in-situ measurements; orange: Model reconstruction.} 
   \end{figure}

\begin{figure}
   \centering
   \subfigure{{\includegraphics[width=\columnwidth]{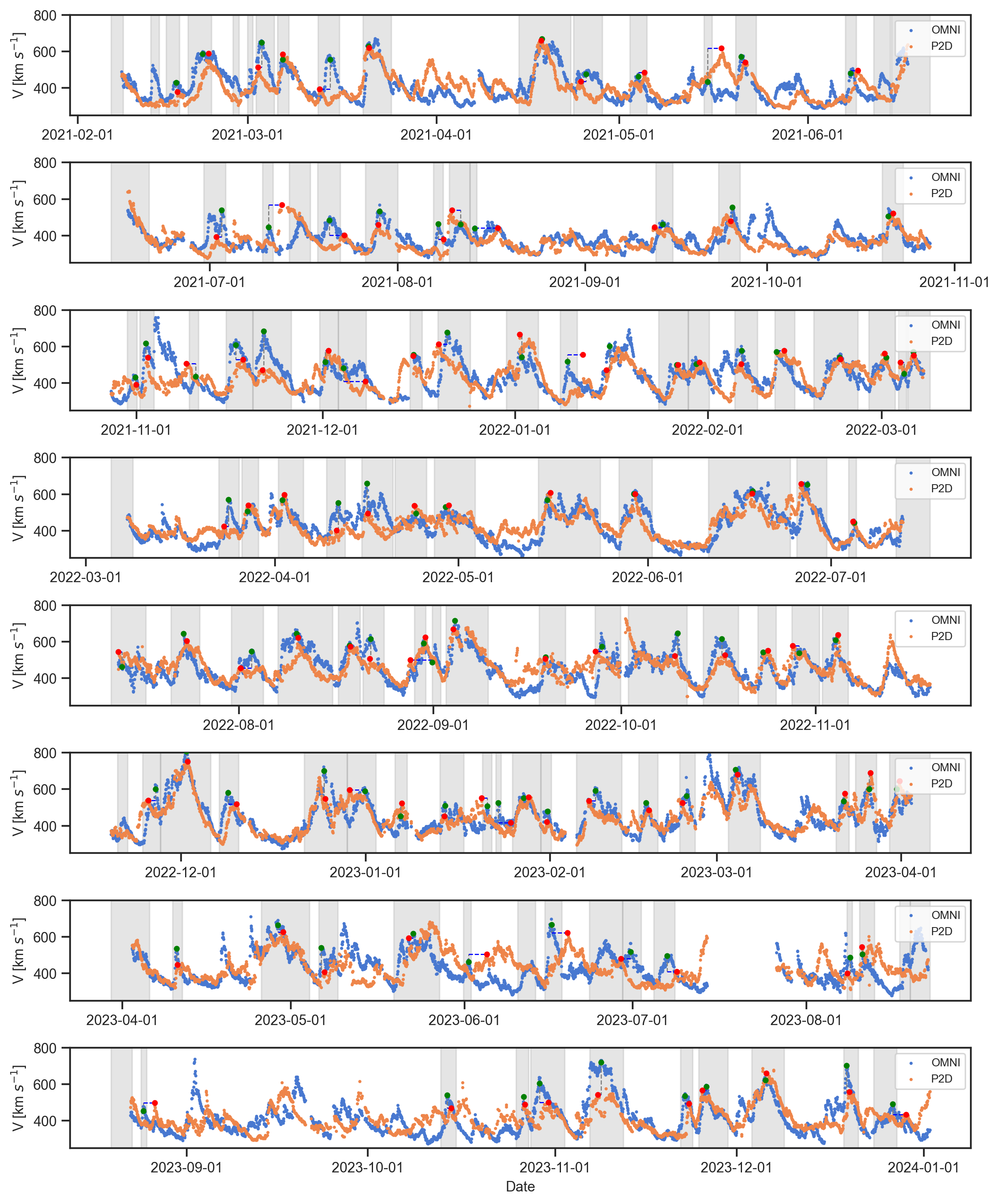}}}

   \caption{\label{fig:Peaks_4}\small Solar wind bulk speed time series with CIR intervals (grey shaded areas) between February 2021 and January 2024. Blue: OMNI in-situ measurements; orange: Model reconstruction.} 
   \end{figure}

Figures \ref{fig:Peaks_1} to \ref{fig:Peaks_4} show the entire bulk speed time series of the P2D reconstruction (orange) along with in-situ measurements from OMNI (blue). Grey shaded areas are CIR times with red and green dots showing their peak velocities from the reconstruction and in-situ data, respectively. 

\begin{figure}
   \centering
   \subfigure{{\includegraphics[width=\columnwidth]{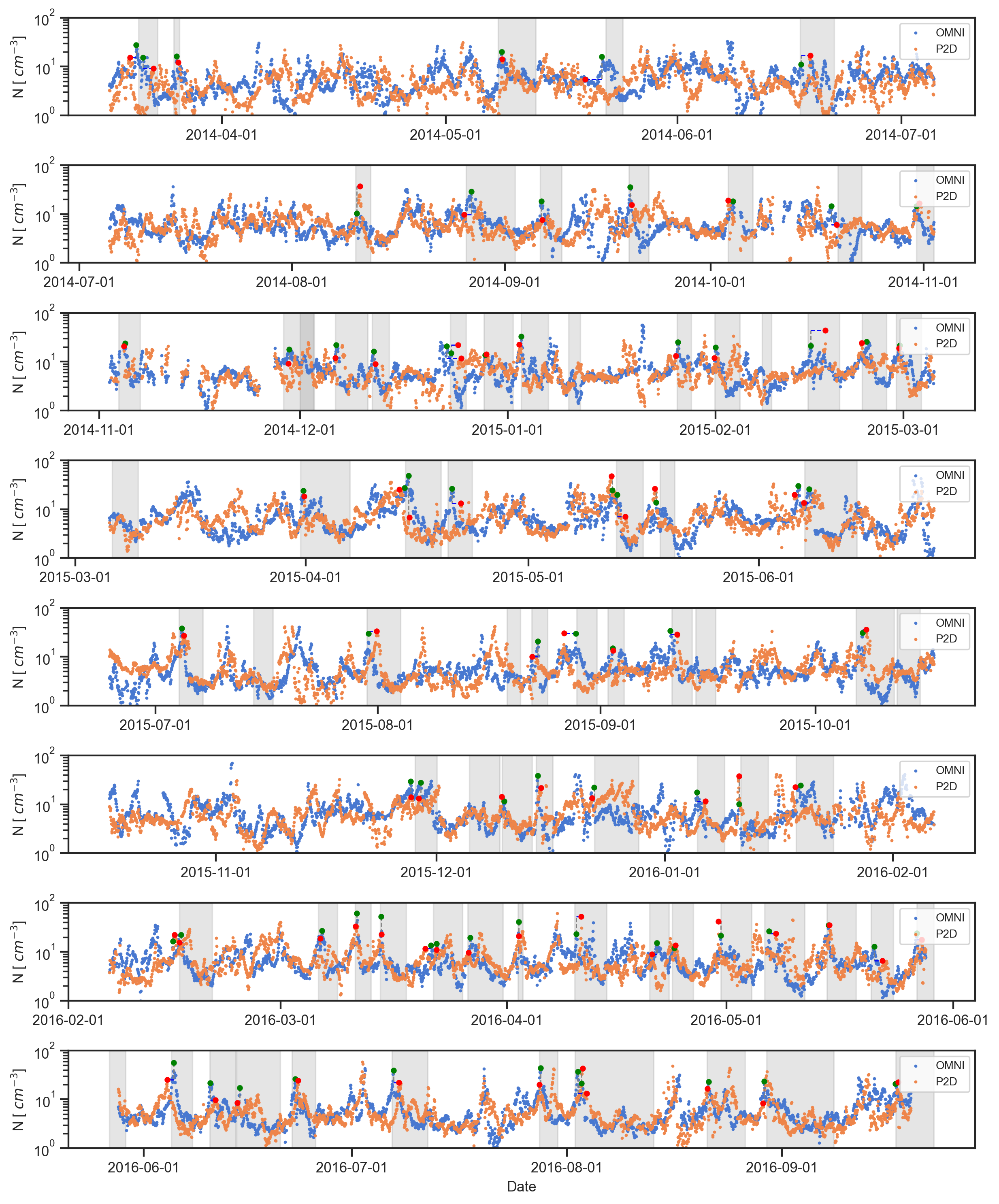}}}

   \caption{\label{fig:Peaks_N1}\small Solar wind proton density time series with CIR intervals (grey shaded areas) between March 2014 and October 2016. Blue: OMNI in-situ measurements; orange: Model reconstruction.} 
   \end{figure}

\begin{figure}
   \centering
   \subfigure{{\includegraphics[width=\columnwidth]{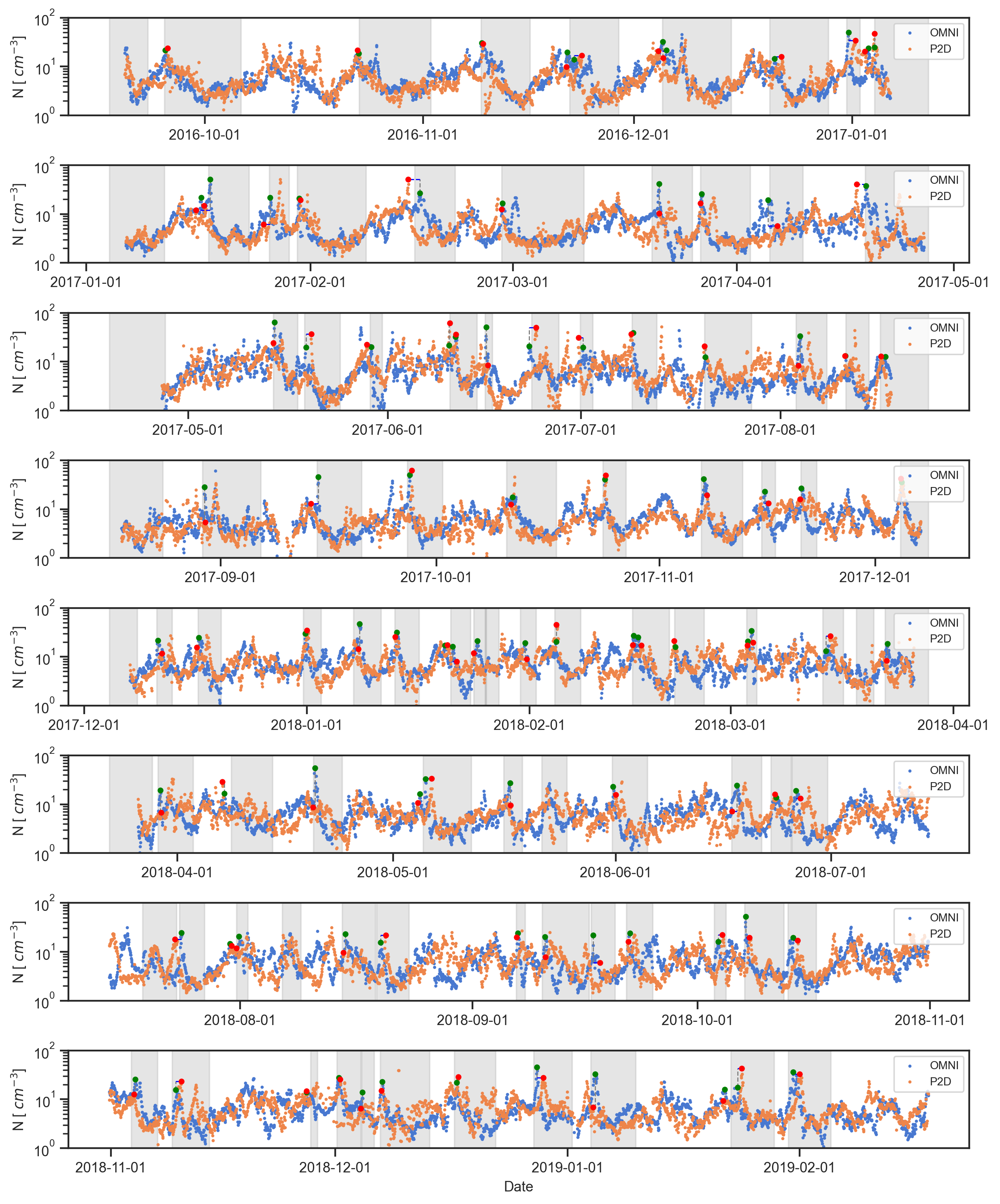}}}

   \caption{\label{fig:Peaks_N2}\small Solar wind proton density time series with CIR intervals (grey shaded areas) between October 2016 and February 2019. Blue: OMNI in-situ measurements; orange: Model reconstruction.} 
   \end{figure}

\begin{figure}
   \centering
   \subfigure{{\includegraphics[width=\columnwidth]{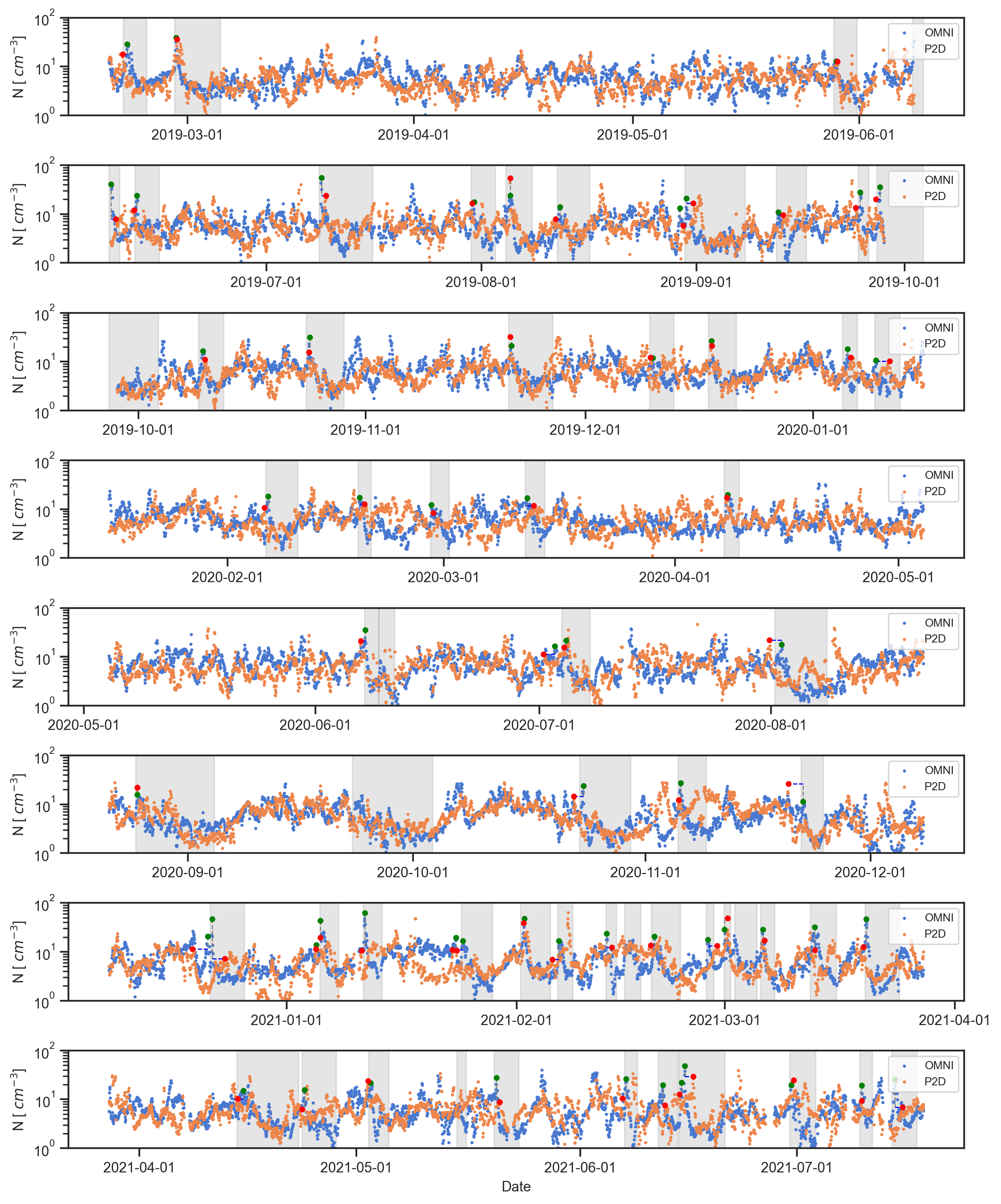}}}

   \caption{\label{fig:Peaks_N3}\small Solar wind proton density time series with CIR intervals (grey shaded areas) between February 2019 and July 2021. Blue: OMNI in-situ measurements; orange: Model reconstruction.} 
   \end{figure}

\begin{figure}
   \centering
   \subfigure{{\includegraphics[width=\columnwidth]{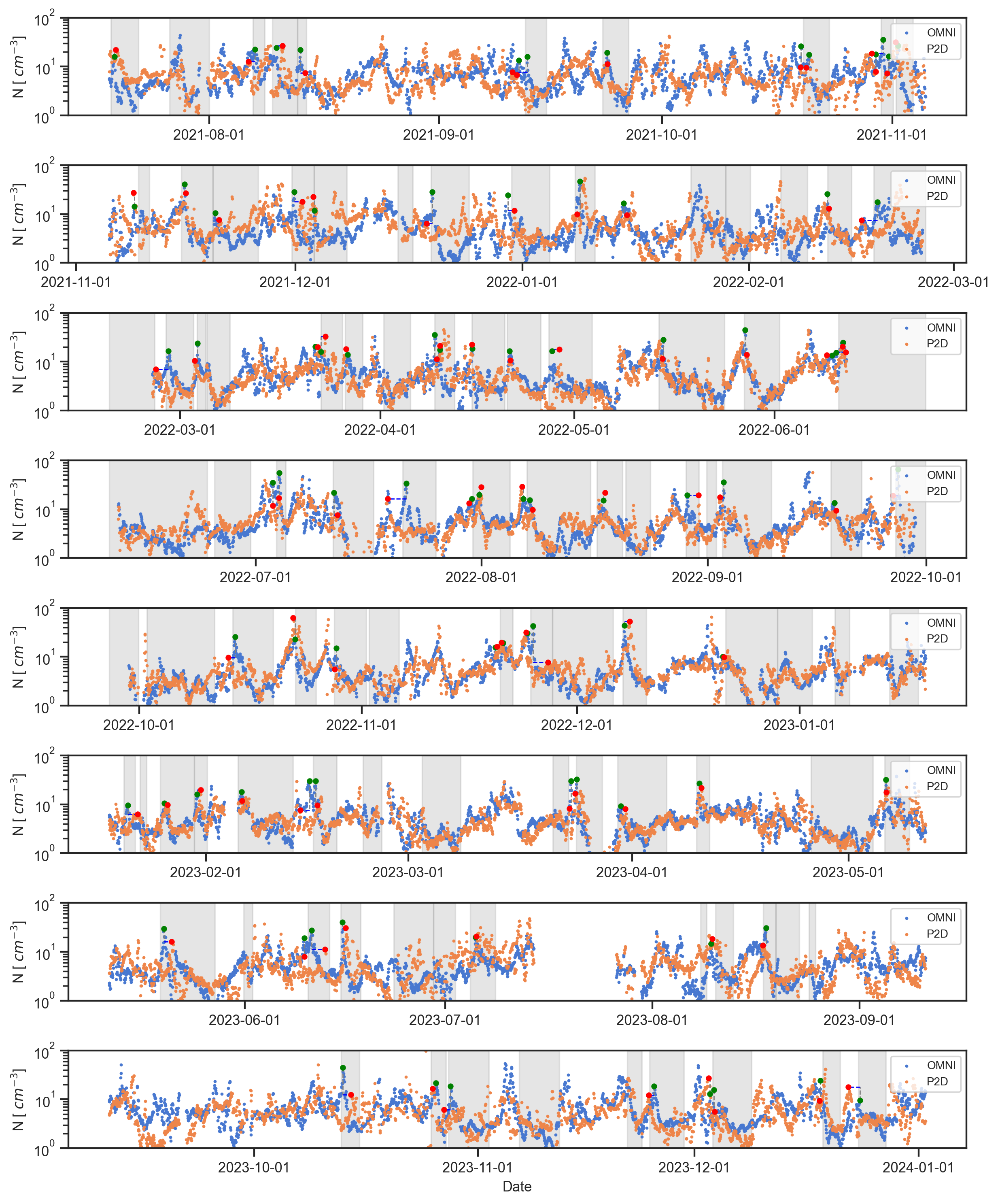}}}

   \caption{\label{fig:Peaks_N4}\small Solar wind proton density time series with CIR intervals (grey shaded areas) between July 2021 and January 2024. Blue: OMNI in-situ measurements; orange: Model reconstruction.} 
   \end{figure}

Figures \ref{fig:Peaks_N1} to \ref{fig:Peaks_N4} show the same concept as Figures \ref{fig:Peaks_1} to \ref{fig:Peaks_4}, but for proton density instead of bulk speed. 

\end{appendix}


\end{document}